\documentstyle[preprint,eqsecnum,aps,amsmath,epsfig]{revtex}
\newcommand{\etal}{{\it et al.}}
\def\expb{\frac{\zeta_i}{1-\zeta_i}}
\def\expbi{\frac{1-\zeta_i}{\zeta_i}}

\begin{document}

\title{\normalsize\rm
       \vspace{+0.3cm} \hfill RIKEN-AF-NP-324, SAGA-HE-143-99  \\
       \vspace{-0.1cm} \hfill KOBE-FHD-99-03, FUT-99-02 \\
                           \ \\
                           \ \\
                           \ \\
\large
       {\bf Polarized Parton Distribution Functions in the Nucleon}
         \ \\ \ \\ }
\normalsize

\author{Y. Goto$^1$, N. Hayashi$^1$, M. Hirai$^2$, H. Horikawa$^3$,\\
        S. Kumano$^2$, M. Miyama$^2$, T. Morii$^3$, N. Saito$^1$,\\
        T.-A. Shibata$^4$, E. Taniguchi$^4$, T. Yamanishi$^5$ \\
        (Asymmetry Analysis Collaboration) \\ \ \\ \ }

\address{{\rm 1~}Radiation Laboratory, The Institute of Physical and
             Chemical Research (RIKEN),\\ Wako, Saitama 351-0198, Japan}
\address{{\rm 2~}Department of Physics, Saga University,
             Saga 840-8502, Japan}
\address{{\rm 3~}Faculty of Human Development, Kobe University,
             Kobe 657-8501, Japan}
\address{{\rm 4~}Department of Physics, Tokyo Institute of Technology,
             Tokyo 152-8551, Japan}
\address{{\rm 5~}Department of Management Science, Fukui University of
             Technology, \\ Gakuen, Fukui 910-8505, Japan \\ \ \\ \ }

\date{December 30, 1999}
\maketitle

\vfill\eject

\begin{abstract}
Polarized parton distribution functions are determined
by using world data from the longitudinally polarized
deep inelastic scattering experiments. A new parametrization
of the parton distribution functions is adopted by taking into
account the positivity and the counting rule. From the fit to
the asymmetry data $A_1$, the polarized distribution functions
of $u$ and $d$ valence quarks, sea quarks, and gluon are obtained.
The results indicate that the quark spin content is $\Delta\Sigma=$0.20
and 0.05 in the leading order (LO) and the next-to-leading-order (NLO)
$\overline{\rm MS}$ scheme, respectively. However,
if $x$ dependence of the sea-quark distribution is fixed
at small $x$ by ``perturbative QCD" and Regge theory, it becomes
$\Delta \Sigma =0.24 \sim 0.28$ in the NLO.
The small-$x$ behavior cannot be uniquely determined by the existing
data, which indicates the importance of future experiments.
From our analysis, we propose one set of LO distributions
and two sets of NLO ones as the longitudinally-polarized
parton distribution functions.
\end{abstract}

\vspace{1.5cm}\hspace{1.05cm}
PACS number(s): 13.60.Hb, 13.88.+e

\vfill\eject


\section{INTRODUCTION}
\label{intro}
For a long time, deep inelastic scattering (DIS) of leptons from the nucleon
has served as an important tool for studying the nucleon substructure
and testing the quantum chromodynamics (QCD). Structure functions
of the nucleon have been measured with this reaction in great precision,
which often provides a firm basis of a search for new physics in hadron
collisions. In addition, basic parameters of QCD such as $\alpha_s$ or
$\Lambda_{\rm QCD}$ have been obtained from the $Q^2$ dependence of
the structure functions. Consequently, hadron-related reactions at high
energies are described by the parton model and perturbative QCD with
reasonable precision.

The measurement of the polarized structure function $g_{1}^{p}(x,Q^2)$
by the European Muon Collaboration (EMC) in 1988 \cite{EMC} has, however,
revealed more profound structure of the proton,
which is often referred to as `{\it the proton spin crisis}'.
Their results are interpreted as very small quark contribution
to the nucleon spin. Then, the rest has to be carried
by the gluon spin and/or by the angular momenta of quarks and gluons.
Another consequence from their measurement was that the strange quark
is negatively polarized, which was not anticipated in a naive quark model.

The progress in the data precision is remarkable in post-EMC experiments.
The final results of the Spin Muon Collaboration (SMC)
experiment \cite{SMC98} have been reported, and its value of $A_1^p$
at the lowest $x$ has decreased in comparison with their previous
one \cite{SMC97}.
The final results of high-precision $A_1^p$ and $A_1^d$ data
have been presented by the Stanford Linear Accelerator Center (SLAC)
E143 collaboration \cite{E14398}, and they consist of more than 200 data
points.
Moreover, the measurement of $g_1^{p}(x,Q^2)$ with
the pure hydrogen target has been carried out by the HERMES
collaboration \cite{HERMES}.
In addition to such improvements in the data precision,
new programs are underway or in preparation at SLAC,
Brookhaven National Laboratory - Relativistic Heavy Ion Collider
(BNL-RHIC)\cite{RHICspin}, European Organization for Nuclear Research
(CERN) \cite{COMPASS}, etc., and results are expected to come out
in the near future.
On the other hand, theoretical advances such as the development of
the next-to-leading-order (NLO) QCD calculations of polarized
splitting functions \cite{MvN,Vogel} stimulated many works on the
QCD analysis of polarized parton distribution functions (PDFs)
\cite{GRSV,GS,ABFR,BBPSS,GGR,SMCfit,LSS}.
There is an attempt to obtain next-to-next-leading order (NNLO)
splitting functions \cite{Gracey} and we can expect further progress
in the precise analysis of polarized PDFs.

In this paper, we present an analysis of world data on
the cross section asymmetry $A_1$ in the polarized DIS processes
for the proton, neutron, and deuteron targets.
We formed a group called Asymmetry Analysis Collaboration (AAC),
and our goal is to determine polarized PDFs,
$\Delta f_i(x,Q^2)$, where $i=u,d,s,\bar{u},\bar{d},\bar{s}, ...,\, \,
{\rm and}\ g$.
Another possible approach is to parametrize structure functions,
$g_1^{N}(x,Q^2)$ ($N=p,n,{\rm and}~d$), which can be expressed
as linear combinations of the PDFs. In the analysis and predictions of
the cross section asymmetry in polarized hadron-hadron collisions,
however, what we need are polarized PDFs rather than structure
functions, because the contribution of each quark flavor is
differently weighted in {\it e.g.} $gq \rightarrow gq$ from 
DIS where each flavor is weighted by electric charge squared.

We choose $A_1$ as the object of the analysis, since it is
more close to the direct observable in experiments than $g_1^{N}(x,Q^2)$.
The $g_1^{N}(x,Q^2)$ data published by the experiments depend
on the knowledge on the unpolarized structure functions at the time
of their publication. By choosing $A_1$ as the object of the analysis,
we can extend the analysis to include new set of data
easily without any change in the previous data set.

As explained in Sec. \ref{parton}, we parametrize the polarized parton
distributions at small momentum transfer squared $Q^2 = 1.0$~GeV$^2$
($\equiv Q_0^2$) with a special emphasis on the positivity and quark
counting rule. Then, they are evolved to the $Q^2$ points, where
the experimental data were taken, by the leading-order (LO)
or NLO $Q^2$ evolution program. Using one of well-established
unpolarized parton distributions, we construct $A_1$ as
\begin{equation}
A_1 (x, Q^2) \simeq \frac{g_1(x,Q^2)}{F_1(x,Q^2)},
\end{equation}
to compare with the experimental data. The polarized parton
distributions at the initial $Q_0^2$ are determined
by a $\chi^2$ analysis.

In Sec. \ref{parton}, we describe the outline of
our analysis with the necessary formulation and the
data set used in the analysis. Section \ref{q2evol} is devoted to
the explanation of the LO and NLO $Q^2$ evolution programs which
we developed for our fit.
The parametrization of the polarized parton distribution
functions at the initial $Q_0^2$ is described in Sec. \ref{paramet},
and the fitting results are discussed in Sec. \ref{results}.
The conclusions are given in Sec. \ref{concl}.


\section{PARTON MODEL ANALYSIS OF POLARIZED DIS DATA}
\label{parton}

In the experiments of polarized DIS, direct observables are the
cross-section asymmetries $A_{\parallel}$ and $A_{\perp}$,
which are defined as
\begin{equation}
A_{\parallel} = \frac{\sigma_{\downarrow \uparrow}-
                      \sigma_{\uparrow \uparrow}}
                     {\sigma_{\downarrow \uparrow}+
                      \sigma_{\uparrow \uparrow}},~~~~~~~~
A_{\perp} = \frac{\sigma_{\downarrow \rightarrow}-
             \sigma_{\uparrow \rightarrow}}
            {\sigma_{\downarrow \rightarrow}+
             \sigma_{\uparrow \rightarrow}}.
\label{E:AsymXS}
\end{equation}
The $\sigma_{\uparrow\uparrow}$ and $\sigma_{\uparrow\downarrow}$
represent the cross sections for the
lepton-nucleon scattering with
their parallel and anti-parallel helicity states, respectively.
On the other hand, the $\sigma_{\uparrow\rightarrow}$ and
$\sigma_{\downarrow\rightarrow}$ are the scattering cross sections for
transversely polarized nucleon target.
We suppress the dependence on $x$ and $Q^2$ where it is evident
hereinafter.
The asymmetries, $A_{\parallel}$ and $A_{\perp}$, are related to the
photon absorption cross section asymmetries, $A_{1}$ and $A_{2}$, by
\begin{equation}
A_{\parallel}={\cal D} (A_{1}+\eta A_{2}),~~~~
A_{\perp}=d(A_{2}-\zeta A_{1}),
\label{E:AsymExp}
\end{equation}
where ${\cal D}$ represents the photon depolarization factor and
$\eta$ is approximated as $\gamma(1-y)/(1-y/2)$ with
$\gamma=2Mx/\sqrt{Q^2}$.
The $d$ and $\zeta$ are other kinematical factors.
The asymmetries, $A_1$ and $A_2$, can be expressed as:
\begin{equation}
A_1(x,Q^2)= \frac{\sigma_{T,\frac{1}{2}}-\sigma_{T,\frac{3}{2}}}
          {\sigma_{T,\frac{1}{2}}+\sigma_{T,\frac{3}{2}}}
   =\frac{g_1(x,Q^2)-\gamma^2 g_2(x,Q^2)}{F_1(x,Q^2)},
\label{E:A1}
\end{equation}
\begin{equation}
A_2(x,Q^2)=\frac{2\sigma_{LT}}{\sigma_{T,\frac{1}{2}}+\sigma_{T,\frac{3}{2}}}
   =\frac{\gamma \, [g_1(x,Q^2)+g_2(x,Q^2) ]}{F_1(x,Q^2)}.
\label{E:A2}
\end{equation}
Here $\sigma_{T,\frac{1}{2}}$ and $\sigma_{T,\frac{3}{2}}$
are the absorption cross sections of virtual transverse photon
for the total helicity of the photon-nucleon system  of $\frac{1}{2}$ and
$\frac{3}{2}$, respectively; $\sigma_{LT}$ is the interference term
between the transverse and longitudinal photon-nucleon amplitudes;
$F_1(x,Q^2)$ is the unpolarized structure function of the nucleon.
If we measure both $A_{\parallel}$ and $A_{\perp}$,  we can extract
both $g_1(x,Q^2)$ and $g_2(x,Q^2)$ from experimental data with
minimal assumptions. Otherwise, $\eta A_2$ should be neglected in
Eq. (\ref{E:AsymExp}) to extract $A_1$. This is justified since $\eta A_2$ is
much smaller than $A_1$ in the present kinematical region.
However, its effect has to be included in the systematic error.
In the small-$x$ or large-$Q^2$ region,
$\gamma^2$ is the order of $10^{-3} - 10^{-2}$.
An absolute value of $g_2(x,Q^2)$ has
been measured to be significantly smaller than $g_1(x,Q^2)$.
Therefore, the asymmetry in Eq. (\ref{E:A1}) can be expressed by
\begin{equation}
A_1(x,Q^2)\simeq \frac{g_{1}(x,Q^2)}{F_{1}(x,Q^2)},
\label{eqn:a1}
\end{equation}
to good approximation. Since the structure function usually
extracted from unpolarized DIS experiments is $F_2(x,Q^2)$,
we use $F_2(x,Q^2)$ instead of
$F_1(x,Q^2)$ by the relation
\begin{equation}
F_1(x,Q^2) = \frac{F_2(x,Q^2)}{2x [ 1+R(x,Q^2) ] }.
\label{eqn:f1}
\end{equation}
The function $R(x,Q^2)$ represents the cross-section ratio
for the longitudinally polarized photon to the transverse one,
$\sigma_{L}/\sigma_{T}$, which is determined experimentally
in reasonably wide $Q^2$ and $x$ ranges in the SLAC
experiment of Ref. \cite{R1990}. Recently published data on $R(x,Q^2)$
by NMC\cite{NMC} showed slightly different values from
the SLAC measurement but mostly agreed within experimental
uncertainties. Therefore, we decided to use SLAC measurements to
be consistent with the most of the analyses of polarized DIS
experiments.

The structure function $F_2$ can be written in terms of
unpolarized PDFs with coefficient functions as
\begin{equation}
F_2(x,Q^2) = \sum\limits_{i=1}^{n_f} e_{i}^2 x
\bigg\{ C_q(x,\alpha_s) \otimes [ q_i(x,Q^2) +\bar{q}_{i}(x,Q^2) ]
      + C_g(x,\alpha_s) \otimes g (x,Q^2)  \bigg\}.
\label{eqn:f2}
\end{equation}
Here $q_i$ and $\bar{q}_i$ are the distributions of quark and antiquark
of flavor $i$ with electric charge $e_i$. The gluon
distribution is represented by $g(x,Q^2)$.
The convolution $\otimes$ is defined by
\begin{equation}
f (x) \otimes g (x) = \int^{1}_{x} \frac{dy}{y}
            f\left(\frac{x}{y} \right) g(y)  .
\end{equation}
The coefficient functions, $C_q$ and $C_g$, are
written as a series in $\alpha_s$ with $x$-dependent coefficients:
\begin{equation}
 C(x,\alpha_s)=\sum\limits_{k=0}^{\infty}
\left (\frac{\alpha_s}{2\pi} \right )^{k} C^{(k)}(x).
\end{equation}
The LO coefficient functions are simply given by
\begin{equation}
C_q^{(0)}(x)=\delta(1-x),~~~~C_g^{(0)}(x)=0.
\end{equation}
In the same way, the polarized structure function $g_1(x,Q^2)$ is
expressed as
\begin{equation}
g_1(x,Q^2) = \frac{1}{2}\sum\limits_{i=1}^{n_f} e_{i}^2
   \bigg\{ \Delta C_q(x,\alpha_s) \otimes [ \Delta q_{i} (x,Q^2)
   + \Delta \bar{q}_{i} (x,Q^2) ] + \Delta C_g(x,\alpha_s) \otimes
   \Delta g (x,Q^2) \bigg \},
\end{equation}
where $\Delta q_i \equiv q^{\uparrow}_i-q^{\downarrow}_i$
($i=u,d,s,...$) represents the difference
between the number densities of quark with helicity parallel
to that of parent nucleon and with helicity anti-parallel.
The definitions of $\Delta \bar q_i$ and $\Delta g$ are the same.
The polarized coefficient functions $\Delta C_q$ and
$\Delta C_g$ are defined similarly to the unpolarized case.

Another separation of the quark distribution can be done by using
flavor-singlet quark distribution $\Delta \Sigma(x,Q^2)$ and
flavor-nonsinglet quark distributions for
the proton and the neutron, $\Delta q_{NS}^{p}(x,Q^2)$ and
$\Delta q_{NS}^{n}(x,Q^2)$, respectively.
Those can be expressed with polarized PDFs as follows:
\begin{eqnarray}
\nonumber
\Delta \Sigma (x) &= & a_0(x) = \Delta u^+ (x) + \Delta d^+ (x)
+ \Delta s^+ (x), \\
\nonumber
\Delta q_{NS}^{p, \, n} (x) & = &\pm \frac{3}{4} \,
                        a_3(x) + \frac{1}{4} \, a_8(x) \\
                            & = & \pm \frac{3}{4} \, 
                    [ \,  \Delta u^+ (x) - \Delta d^+ (x) \, ]
 + \frac{1}{4} \, [ \, \Delta u^+ (x) + \Delta d^+ (x) 
                            - 2 \Delta s^+ (x) \, ] ,
\end{eqnarray}
where $\Delta u^+ (x) = \Delta u(x) + \Delta \bar{u}(x)$ and
similarly for $\Delta d^+ (x)$ and $\Delta s^+ (x)$. 
Analyses in Ref. \cite{ABFR} and Ref. \cite{SMCfit} utilized
this separation. Such separation is useful in $Q^2$ evolution, and
it is also natural when one wants to obtain quark
contribution to the proton spin, $\int_0^1 \Delta \Sigma (x) dx$.

On the other hand, when we try to calculate the cross section
for polarized $pp$ reaction, {\it e.g.} Drell-Yan production
of lepton pairs, we need the combination of
$\Delta q_{i}(x_1 ) \times \Delta \bar{q}_{i}(x_2)$
(multiplied by electric charge squared). To allow such
calculations with the above separation,
we need further assumption on the polarized antiquark
distributions, {\it e.g.} flavor symmetric sea,
$\Delta u_{\rm sea}(x) = \Delta \bar{u}(x) =
\Delta d_{\rm sea}(x) = \Delta \bar{d}(x) = \Delta s(x) =
\Delta \bar{s}(x)$. With such assumption, the above separation
becomes equivalent to the PDF separation in a sense that one
description can be translated to another by simple transformation.

Of course, we already know that unpolarized sea-quark distributions
are not flavor symmetric \cite{sk} from various experiments including
Drell-Yan production of lepton pairs in $pp$ and $pd$ collisions.
Therefore, this assumption is only justified
as an approximation due to limited experimental data.
In principle, charged-hadron production data could clarify
this issue. Although a $\chi^2$ analysis for the SMC
and HERMES data seems to suggest a slight $\Delta\bar u$ excess
over $\Delta \bar d$ \cite{my}, the present data are not accurate
enough for finding such a flavor asymmetric signature.
Future experiments with charged current at RHIC \cite{w}
and polarized option at HERA will be very useful in improving
our knowledge on the spin-flavor structure of the nucleon.
Furthermore, as it has been done in the unpolarized studies,
the difference between the polarized $pp$ and $pd$ cross sections
provides a clue for the polarized flavor asymmetry \cite{km} although
actual experimental possibility is uncertain at this stage.

The parametrization models studied so far have various differences
in other aspects:
(a) the choice of the renormalization scheme, (b) the functional form of
the polarized parton distributions due to different physical requirements
at $Q_0^2$, and (c) the physical quantity to be fitted.
In the following, we describe our position on these issues.

\begin{itemize}
\item {\bf Renormalization Scheme} \\
Although the parton distributions have no scheme
dependence in the LO, they do depend on the renormalization scheme
in the NLO and beyond. In the polarized case, we have different choices
of the scheme due to the axial anomaly and the ambiguity in treating
the $\gamma_5$ in $n$ dimensions\cite{LSS}. In the NLO analysis,
the widely-used scheme is the modified minimal subtraction
($\overline {\rm MS}$) scheme, in which the first moment of
the nonsinglet distribution is $Q^2$-independent.
It was used, for example, by Mertig and van Neerven \cite{MvN}
and Vogelsang \cite{Vogel}. However, the first moment of
the singlet distribution is $Q^2$-dependent in this scheme
and thus it is rather difficult to compare the value of
$\Delta \Sigma(x,Q^2)$
extracted from the DIS at large $Q^2$ with the one from the
static quark model at small $Q^2$.  To cure this difficulty,
Ball, Forte and Ridolfi \cite{Ball} used the so-called
AB (Adler-Bardeen) scheme, in which the first moment of the singlet
distribution becomes independent of $Q^2$ because
of the Adler-Bardeen theorem\cite{Adler}.
In those schemes, however, some soft contributions are included in
the Wilson coefficient functions and not completely absorbed
into the PDFs. Another scheme called
the JET scheme \cite{AFL} or the
CI (chirally invariant) scheme \cite{Cheng} has been recently proposed.
All the hard effects are absorbed into the Wilson coefficient
functions in this scheme.

Although we choose the $\overline{\rm MS}$ scheme in our analysis,
the polarized PDFs in one scheme are related to those in other schemes
with simple formulae \cite{LSS}.

\item{\bf Functional Form of polarized PDF and Physical Requirements}\\
Different functional forms have been proposed so far
for the polarized PDFs by taking account of various physical conditions.
We choose the functional form with the special emphasis
on the positivity condition and quark counting rule \cite{Counting}
at $Q^2_0=1.0$~GeV$^2$.

The positivity condition is originated in a probabilistic
interpretation of the parton densities. The polarized PDFs
should satisfy the condition
\begin{equation}
| \, \Delta f_i(x, Q^2_0) \, | \leq f_i(x, Q^2_0) .
\label{eqn:POSCON}
\end{equation}
This is valid in the LO since we can have the complete probabilistic
interpretation for each polarized distribution only at the LO.
Even in NLO, however, the positivity condition for the polarized
cross section $\Delta \sigma$ with the unpolarized cross section
$\sigma$,
\begin{equation}
|\,\Delta \sigma \, | \leq \sigma,
\label{eqn:cross-pos}
\end{equation}
should still apply for any processes to be calculated with the
polarized PDFs to the order of ${\cal O}(\alpha_s)$.
Since it is very difficult to calculate the polarized and
unpolarized cross sections of the NLO for all the possible
processes, it is not realistic to determine the polarized NLO
distributions by the positivity condition of Eq. (\ref{eqn:cross-pos}).
In our analysis, we simply require that Eq. (\ref{eqn:POSCON})
should be satisfied in the LO and also NLO at $Q_0^2$.
It is shown in Ref. \cite{LeaderPositivity}
that the NLO $Q^2$ evolution should preserve the positivity
maintained at initial $Q^2_0$.

In many cases, Regge behavior has been assumed for $x\rightarrow 0$, 
and the color coherence of gluon couplings has been also used
at $x\simeq 0$ \cite{BBS}.
Furthermore, it is an interesting guiding principle that
the polarized distributions have a similar behavior
to the unpolarized ones in the large-$x$ region \cite{LSS}.
Since behavior of the distributions at large $x$
is determined by the term $(1-x)^{\beta}$ in the functions,
where $\beta$
is a constant, we simply require that the polarized distributions
should have the same $(1-x)^{\beta}$ term as the unpolarized ones.

Those physical requirements and assumptions have to be tested by
comparing with the existing experimental data.

As for the choice of $Q_0^2$, it has to be large enough
to apply perturbative QCD, but it should be small enough to maintain
a large set of experimental data. We find $Q_0^2$=1.0~GeV$^2$ to
be a reasonable choice in our analysis.

\item{\bf Physical Quantities to be Fitted}\\
In most of the polarized experiments, the data have been presented
for $A_1(x, Q^2)$ and $g_1(x, Q^2)$. Some
analyses~\cite{ABFR,GGR,SMCfit,BBS}
used the $g_1(x, Q^2)$ as data samples, while others
\cite{GRSV,GS,BBPSS,LSS} used the $A_1(x, Q^2)$.
It should be, however, noted that $g_1(x, Q^2)$ is obtained by
multiplying $A_1(x, Q^2)$ by $F_1(x, Q^2)$, so that
it is not free from ambiguity of the
unpolarized structure function, $F_1(x, Q^2)$.
Therefore, we consider that it is more advantageous to use the
$A_1(x, Q^2)$ as the data samples not only for the current work
but also for the convenience in expanding the data set to include
new data set from SLAC, DESY (German Electron Synchrotron),
CERN, and RHIC.

Another important quantity which we should carefully
consider is the cross section ratio $R(x,Q^2)=\sigma_L/\sigma_T$, where
$\sigma_L$ and $\sigma_T$ are absorption cross sections of
longitudinal and transverse photons, respectively.
In principle, nonzero $R(x,Q^2)$ is originated from radiative corrections
in perturbative QCD, higher twist effects, and target mass effects.
Higher twist contribution to $R(x,Q^2)$ is expected to be small
in the large $Q^2$ region.
So far, some analyses employed nonzero $R(x,Q^2)$, while other
analyses assumed $R(x,Q^2)=0$. However, the latter is not consistent with
the experimental analysis procedure, since $R(x, Q^2)$ is also
used for the evaluation of photon depolarization factor ${\cal D}$.
Indeed our analysis shows that world data prefer $R(x,Q^2) \neq 0 $:
the $\chi^2$ increases significantly with $R=0$.
Therefore, we use nonzero $R(x,Q^2)$ in fitting the data of $A_1(x, Q^2)$.
\end{itemize}


Table \ref{T:Exp} summarizes experiments with published data on
the polarized DIS
\cite{EMC,SMC98,SMC97,E14398,HERMES,E130,E142,E154,E155}.
These measurements cover a wide range of $x$ and
$Q^2$ with various beam species and energies and various
types of polarized nucleon target (not shown in the table).
The listed are the number of data points above $Q^2 =1.0$~GeV$^2$,
and the total number of data points are 375.

We use the data with minimal manipulation to analyze them
in our framework so as to be consistent with
the $Q^2$ evolution, the unpolarized parton distributions,
and the function $R(x,Q^2)$.
For example, the E143 provides the proton data which are obtained by
combining the results of different beam energies
using the weights based on the unpolarized cross sections \cite{E14398}
(28 points), in addition to ``{\it raw}" data for each beam energy
(81 points at $Q^2 >$1~GeV$^2$). Such weights depend on the choice of the
unpolarized structure functions, which are being updated.
To localize dependence on the unpolarized structure functions in the final
manipulation for getting $g_1(x,Q^2)$, {\it i.e.}
$A_1(x,Q^2)$ multiplied by $F_1(x,Q^2)$,
we decided to use the ``{\it raw}" data in our analysis.

Table \ref{T:Exp} also includes analysis methods.
One of the major differences in the analysis is the treatment
of the $A_2(x,Q^2)$ and $g_2(x,Q^2)$ contributions to the
$g_1(x,Q^2)/F_1(x,Q^2)$. Some of SLAC experiments measured
both $A_{\parallel}$ and $A_{\perp}$ to enable direct extraction of
$g_1/F_1$ and $g_2/F_1$. Other experiments included possible contribution
of $\eta A_2$ in their estimation of systematic errors.

As mentioned above, the choice of the function
$R(x,Q^2)$ potentially affects $A_1(x,Q^2)$, thus final results on
polarized PDFs, since the function
affects the photon depolarization
factor ${\cal D}$.
While it was assumed to be constant in the analyses of
the early days, its $x$-dependence and $Q^2$-dependence have been
found to be significant \cite{R1990}. To reflect the most updated
knowledge of $R(x,Q^2)$ on our analysis, we have reevaluated
the E130 and EMC data by using $R_{\rm 1990}(x,Q^2)$ \cite{R1990},
which most of the experiments employed.
However, we found changes of a few percent in EMC data and about 10\%
in E130 data: both of them are smaller than experimental errors.


\section{Q$^2$ EVOLUTION}
\label{q2evol}
In our framework and in most of the analyses of structure functions in
the parton model, the polarized parton distributions are provided at certain
$Q^2 (= Q_0^{\, 2})$ with a number of parameters, which are determined
so as to fit polarized experimental data.
The experimental data, in general, range over a wide $Q^2$ region.
The polarized parton distributions have to be evolved from $Q_{0}^{\,2}$ to
the $Q^2$ points, where experimental data were obtained,
in the $\chi^2$ analysis.
In calculating the distribution variation from
$Q_0^{\, 2}$ to given $Q^2$,
the Dokshitzer-Gribov-Lipatov-Altarelli-Parisi (DGLAP) evolution equations
are used.
%

To compare our parametrization with the data, we need to construct $A_1(x,Q^2)$
from the polarized and unpolarized PDFs.
Since the determination of the unpolarized PDFs
is not in our main scope,
we decided to employ one of the widely-used set of PDFs.
Although there are slight variations among the unpolarized parametrizations,
the calculated $F_2(x,Q^2)$ structure functions are essentially the same
because almost the same set of experimental data is used in
the unpolarized analyses.
The Gl\"uck-Reya-Vogt (GRV) unpolarized distributions
\cite{GRV98} have been used in our analyses; however,
the parametrization results do not change significantly
even with other unpolarized distributions.
We checked this point by comparing the GRV $F_2(x,Q^2)$ structure function
with those of MRST (Martin-Roberts-Stirling-Thorne) \cite{MRST} and 
CTEQ (Coordinated Theoretical/Experimental Project on QCD 
Phenomenology and Tests of the Standard Model) \cite{CTEQ}
at $Q^2$=5~GeV$^2$ in the $x$ range $0.001 < x < 0.7$.
The differences between these distributions are merely less than about 3\%.
The differences depend on the $x$ region; however,
we find no significant systematic deviation from the GRV distribution.

We calculate the GRV unpolarized distributions
at $Q_0^{\, 2}$=1 GeV$^2$ in Ref.\cite{GRV98}
\footnote{Actual calculation has been done by the FORTRAN program,
which was obtained from the www site,
http://durpdg.dur.ac.uk/HEPDATA/PDF.}.
The distributions are evolved to those at $Q^2$ by the DGLAP
equations, then they are convoluted with the coefficient functions
by Eq. (\ref{eqn:f2}). Because the unpolarized evolution equations
are essentially the same as the longitudinally polarized ones
in the following, except for the splitting functions,
we do not discuss them in this paper. The interested reader may read,
for example, Ref. \cite{q2-evol}.

The polarized PDFs are provided at the initial $Q_0^{\, 2}$;
therefore, they should be evolved to $Q^2$ by the DGLAP equation
in order to obtain $g_1(x,Q^2)$.
The DGLAP equations are coupled integrodifferential
equations with complicated splitting functions in the NLO case.
Both the LO and NLO cases can be handled by the same
DGLAP equation form; however, the NLO effects are included in the running
coupling constant $\alpha _s (Q^2)$ and in the splitting functions
$\Delta P_{ij} (x)$.

In solving the evolution equations,
it is more convenient to use the variable $t$ defined by
\begin{equation}
t \equiv \ln Q^2,
\label{t}
\end{equation}
instead of the variable $Q^2$.
Then, the flavor nonsinglet DGLAP equation is given by
\begin{equation}
\frac{\partial}{\partial t}  \Delta q_{_{NS}}(x,t) =
\frac{\alpha_s (t)}{2\pi}
\Delta P_{q^\pm, {NS}} (x) \otimes \Delta q_{_{NS}}(x,t) ,
\label{eqn:ns}
\end{equation}
where $\Delta q_{_{NS}}(x,t)$ is a longitudinally-polarized nonsinglet
parton distribution, and $\Delta P_{q^\pm, {NS}}$ is the polarized
nonsinglet splitting function. The notation $q^\pm$
in the splitting function indicates
a ``$\Delta q \pm \Delta \bar q$ type" distribution
$\sum_i a_i (\Delta q_i \pm \Delta \bar q_i)$, where
$a_i$ is given constant with flavor $i$.
The singlet evolution is more complicated than the nonsinglet one
due to gluon participation in the evolution. The singlet quark distribution
is defined by
$\Delta \Sigma (x,t) = \sum_i^{N_{f}} (\Delta q_i+\Delta \bar{q}_i )$,
and its evolution is described by the coupled integrodifferential
equations,
\begin{equation}
\frac{\partial}{\partial t}
\left(\begin{array}{c}
  \Delta \Sigma(x,t) \\
  \Delta g (x,t)
\end{array} \right) = \frac{\alpha_s (t)}{2\pi} \,
\left( \begin{array}{cc}
  \Delta P_{qq}(x) & \Delta P_{qg}(x) \\
  \Delta P_{gq}(x) & \Delta P_{gg}(x) \\
\end{array} \right)  \otimes
\left( \begin{array}{c}
  \Delta \Sigma(x,t) \\
  \Delta g(x,t)
\end{array} \right) .
\label{eqn:sing}
\end{equation}

The numerical solution of these integrodifferential equations is
obtained by a so-called brute-force method.
The variables $t$ and $x$ are divided into small steps, $\delta t_i$ and
$\delta x_i$ respectively, and
then the integration and differentiation are defined by
\begin{align}
\frac{d f(x)}{dx} &= \frac{f(x_{m+1})-f(x_m)}{\delta x_m} ,\\
\int f(x)\ dx &= \sum_{m=1}^{N_x} \delta x_m \, f(x_m) .
\end{align}
The evolution equation can be solved numerically with
these replacements in the DGLAP equations.
This method seems to be too simple; however, it has an advantage
over others not only in computing time but also in future applications.
For example, the evolution equations with higher-twist effects cannot
be solved by orthogonal polynomial methods.
It is solved rather easily by the brute-force method \cite{q2-evol}.
Another popular method is to solve the equations in the moment space.
However, the $x$ distributions are first transformed into the
corresponding moments. Then, the evolutions are numerically solved.
Finally, the evolved moments are again transformed into the $x$ 
distributions. If the distributions are simple enough to be handled
analytically in the Mellin transformation, it is a useful method.
However, if the distributions become complicated functions in future
or if they are given numerically, errors may accumulate in the numerical
Mellin and inverse Mellin transformations.
Therefore, our method is expected to provide potentially better
numerical solution although it is very simple. 

The employed method is identical to that in Ref. \cite{q2-evol}
in its concept, but we had to improve the program in its computing time,
since the evolution subroutine is called a few thousand times in searching
for the optimum set of polarized distributions.
There are two major modifications. The first one is to change
the method of the convolution integrals, and the second is to
introduce the cubic Spline interpolation for obtaining
the parton distributions during the evolution calculation.
Previously we calculated the convolution integral by
$\int_x^1 \frac{dy}{y} \, \Delta P (x/y) \Delta q(y, t)$. In this
case,  we had to calculate the splitting functions for each $x$
value in the numerical integration, since the integration variable
and the argument of the splitting function are different.
Because the NLO splitting functions are complicated,
this part of calculation consumed much time. In the present program,
we evaluate the integral by $\int_x^1 \frac{dy}{y} \,
\Delta P (y) \Delta q(x/y, t)$, which is mathematically
equivalent to the above integral, and thus, we only need
to calculate the splitting functions at a fixed set of $x$ values once
before the actual evolution.
For example, the nonsinglet equation, Eq. (\ref{eqn:ns}),
becomes
\begin{equation}
\Delta q_{_{NS}}(x_k,t_{j+1}) = \Delta q_{_{NS}}(x_k,t_j)
   + \delta t_j  \, \frac{\alpha_s (t)}{2\pi} \,
    \sum_{m=k}^{N_x} \frac{\delta x_m}{x_m}  \,
    \Delta P_{q^\pm, NS}  ( x_m )            \,
    \Delta q_{_{NS}} \left( \frac{x_k}{x_m},t_j \right) .
\label{eqn:bfns}
\end{equation}
If the initial distribution $\Delta q_{_{NS}}(x_k/x_m,t_0=0)$ is provided,
the next distribution $\Delta q_{_{NS}}(x_k,t_1)$ is calculated
by the above equation. Then, $\Delta q_{_{NS}}(x_k/x_m,t_1)$ is
calculated by the cubic Spline interpolation.
Repeating this step $N_t-1$ times, we obtain
the evolved nonsinglet distribution $\Delta q_{_{NS}}(x_k,t_{N_t})$.
With these refinements, the evolution equations are solved significantly
faster, and the subroutine can be used in the parametrization study.

\begin{figure}[h]
\addcontentsline{toc}{section}{\protect\numberline{\S}{FIGURES}}
      \vspace{0.0cm}
   \begin{center}
       \epsfig{file=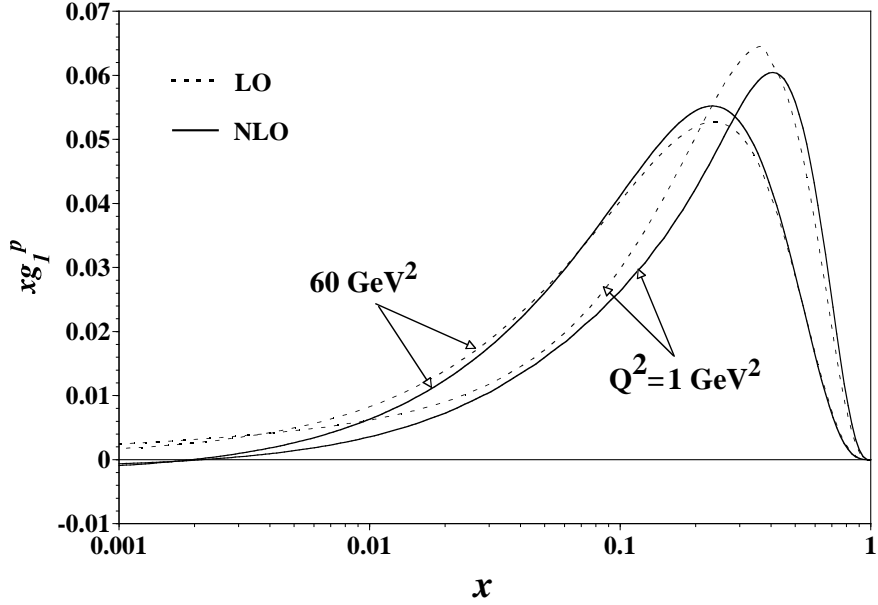,width=12.0cm}
   \end{center}
   \vspace{-0.2cm}
   \caption{$Q^2$ evolution results for the proton structure function
          $g_1^{\, p}$. The initial LO and NLO-1 $g_1$ structure functions
          are evolved to those at 60 GeV$^2$ by the LO and NLO DGLAP
          evolution equations.}
   \label{fig:xg1}
\end{figure}

We show the $Q^2$ dependence in $g_1^p (x,Q^2)$ and $A_1^p (x,Q^2)$
as a demonstration of the performance of our program.
The numerical calculations are done such that the accuracy becomes
better than about 2\% in the asymmetry $A_1^p$. 
The LO and NLO (set NLO-1) parton distributions obtained in our
analyses are used. The details of these distributions are discussed
in Sec. \ref{results}. The initial structure functions $g_1$
at $Q^2=1.0$ GeV$^2$ are evolved to those at $Q^2=60.0$ GeV$^2$.
Most of the used $A_1$ data are within this $Q^2$ range.
The LO and NLO results are shown in Fig. \ref{fig:xg1} by the dashed
and solid curves, respectively. The LO distributions tend to be
shifted to the smaller $x$ region than the NLO ones.
There are two reasons for the differences between the LO and NLO
distributions. One is the difference between the LO and
NLO $F_2$ structure functions for fitting the same data set of $A_1$,
and the other is the difference in $Q^2$ evolution.

\begin{figure}[h]
      \vspace{0.0cm}
   \begin{center}
       \epsfig{file=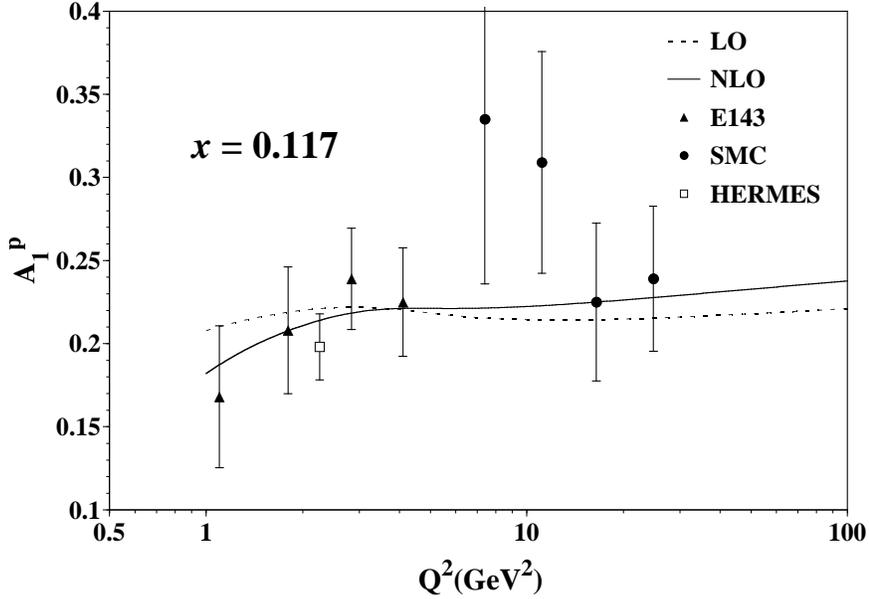,width=12.0cm}
   \end{center}
   \vspace{-0.2cm}
 \caption{Calculated LO and NLO spin asymmetries $A_1$ for the proton are
          compared with the experimental results by the SMC,
          SLAC-E143, and HERMES collaborations at $x \approx 0.117$.
          The theoretical curves are obtained by using our LO and NLO-1
          fitting results at $Q^2$=1 GeV$^2$.}
 \label{fig:asym-1}
\end{figure}

In Fig. \ref{fig:asym-1}, our $Q^2$ evolution curves at $x=0.117$ are
shown with the asymmetry $A_1$ data by the SMC \cite{SMC98},
SLAC-E143 \cite{E14398}, and HERMES \cite{HERMES} collaborations.
The initial distributions are our LO and NLO parametrizations
at $Q^2=1$ GeV$^2$. The dashed and solid curves indicate
the LO and NLO evolution results, respectively. In the large $Q^2$
region, both $Q^2$ variations
($\partial A_1/ \partial \ln Q^2$) are almost the same;
however, they differ significantly at small $Q^2$, particularly
in the region $Q^2<\, $2 GeV$^2$. As the $Q^2$ becomes smaller, the NLO
contributions become more apparent. We find that the theoretical asymmetry
has $Q^2$ dependence although it is not large at $x=0.117$.
It is often assumed that the experimental asymmetry $A_1(x,Q^2)$
is independent of $Q^2$ by neglecting the $Q^2$ evolution difference
between $g_1(x,Q^2)$ and $F_1(x,Q^2)$ in extracting the
$g_1(x,Q^2)$ structure functions.
The assumption has no physical basis.
For a precise analysis, the $Q^2$ dependence in the asymmetry
has to be taken into account properly and our framework is ready for
such precision studies.


\section{PARAMETRIZATION OF POLARIZED PARTON DISTRIBUTIONS}
\label{paramet}

Now, we explain how the polarized parton distributions are parametrized.
The unpolarized PDFs $f_i(x, Q^2_0)$ and polarized PDFs
$\Delta f_i(x, Q^2_0)$ are given at the initial scale $Q^2_0$.
Here, the subscript $i$ represents quark flavors and gluon.
These functions are generally assumed to be in a factorized form
of a power of $x$ inspired by Regge-like behavior at small $x$,
a polynomial of $x$ at medium $x$,
and a power of $(1-x)$ expected from the counting rule at large $x$:
{
\setcounter{enumi}{\value{equation}}
\addtocounter{enumi}{1}
\setcounter{equation}{0}
\renewcommand{\theequation}{\arabic{section}.\theenumi\alph{equation}}
\begin{eqnarray}
f_i(x, Q^2_0) & = & C_i \, x^{\alpha_{1 i}} \,
           (1 - x )^{\alpha_{2 i}} \, (1 + \sum_j \alpha_{3 i,j} \,
           x^{\alpha_{4 i,j}}) , \\
\label{eqn:PDFEQ1}
\Delta f_i(x, Q^2_0) & = & D_i~x^{\beta_{1 i}} \,
           (1 - x )^{\beta_{2 i}} \, (1 + \sum_j \beta_{3 i,j} \,
           x^{\beta_{4 i,j}}) ,
\label{eqn:PDFEQ2}
\end{eqnarray}
\setcounter{equation}{\value{enumi}}}
where $C_i$ and $D_i$ are normalization factors and
$\alpha_{1 i}$, $\alpha_{2 i}$, $\alpha_{3 i,j}$, $\alpha_{4 i,j}$,
$\beta_{1 i}$, $\beta_{2 i}$, $\beta_{3 i,j}$, and $\beta_{4 i,j}$ are
free parameters.

From the best fit to all the experimental data of
the polarized DIS including
new data, we can determine, in principle,
the parameters in Eq. (\ref{eqn:PDFEQ2}).
In practice, however, some of the parameters highly correlate each other
and it is difficult to determine all the parameters independently.
Therefore, it is desirable to reduce the number of parameters by
applying physical conditions instead of leaving all these parameters
free.

In the present analysis, to constrain the explicit forms of
polarized PDFs,
we require two natural conditions: (i) the positivity condition of
the PDFs and (ii) the counting rule for
the helicity-dependent parton distribution functions.

In order to make the positivity condition of Eq. (\ref{eqn:POSCON}) be
tractable in the numerical analysis, we modify the functional form of
the polarized PDF as
\begin{equation}
\Delta f_i(x, Q^2_0) = h_i (x) \, f_i(x, Q^2_0),
\label{eqn:PPDF1}
\end{equation}
where
\begin{equation}
h_i (x) = A_i \, x^{\alpha_i} \,
(1 - x )^{\beta_i} \, (1 + \gamma_i \, x^{\lambda_i}),
\label{eqn:gx1}
\end{equation}
at the initial scale $Q^2_0$.
Therefore, the positivity condition can be written as
\begin{equation}
| \, h_i (x) \, | \le 1
\label{eqn:hless1}
\end{equation}

Furthermore, taking into account of the counting rule mentioned
in Section II, we reduce Eq. (\ref{eqn:gx1}) to
\begin{equation}
h_i (x) = A_i \, x^{\alpha_i} \, (1 + \gamma_i \, x^{\lambda_i}) ,
\label{eqn:gx2}
\end{equation}
and we have the following functional form of polarized PDFs at $Q^2_0$:
\begin{equation}
\Delta f_i(x, Q^2_0) = A_i \, x^{\alpha_i} \,
               (1 + \gamma_i \, x^{\lambda_i}) \, f_i(x, Q^2_0) .
\label{eqn:PPDF2}
\end{equation}
Thus, we have four parameters ($A_i$, $\alpha_i$, $\gamma_i$ and
$\lambda_i$) for each $i$.

We further reduce the number of free parameters
by assuming the SU(3) flavor symmetry for the sea-quark
distributions at $Q_0^2$.
As mentioned in Section II, this is simply a compromise
due to a lack of experimental data.
It should be noted that the sea-quark distributions are not
SU(3) flavor symmetric at $Q^2 > Q_0^2$ even
with the symmetric distributions at the initial $Q_0^2$.

When we assume this SU(3) flavor symmetric sea,
the first moments of $\Delta u_v (x)$ and $\Delta d_v (x)$
for the LO, which are written as $\eta_{u_v}$ and $\eta_{d_v}$,
respectively, can be described in terms of
axial charges for octet baryon,
$F$ and $D$ measured in hyperon and neutron $\beta$-decays
as follows:
\begin{eqnarray}
\eta_{u_v}-\eta_{d_v}&=&F + D,
\nonumber \\
\eta_{u_v}+\eta_{d_v}&=&3F - D.
\label{eqn:UVDV}
\end{eqnarray}
Note that Eq. (\ref{eqn:UVDV}) is also used for the NLO
($\overline{{\rm MS}}$) case.
Recently, since the $\beta$-decay constants
have been updated \cite{PDG98},
we reevaluate $F$ and $D$ from the $\chi^2$ fit to the
experimental data of four different semi-leptonic decays:
$n \rightarrow p$, $\Lambda \rightarrow p$, $\Xi \rightarrow \Lambda$,
and $\Sigma \rightarrow n$, by assuming the SU(3)$_f$ symmetry
for the axial charges of octet baryon.
With $\chi^2$/d.o.f=0.98, the $F$ and $D$ are determined as
\begin{eqnarray}
& &F = 0.463\pm0.008,
\nonumber \\
& &D = 0.804\pm0.008,
\label{eqn:FandD}
\end{eqnarray}
which lead to $\eta_{u_v}=0.926\pm 0.014$ and
$\eta_{d_v}=-0.341 \pm 0.018$. 
In this way, we fix these two moments at their central values,
so that two parameters $A_{u_v}$ and $A_{d_v}$ are determined
by these first moments and other parameter values. 
Thus, the remaining job is to determine the values of remaining
14 parameters,
$A_{\bar q}, \ A_g, \ \alpha_i, \ \gamma_i, \ \lambda_i 
\ (i=u_v, \, d_v, \, \bar q, \, g)$,
by a $\chi^2$ analysis of the polarized DIS experimental data.


\section{NUMERICAL ANALYSIS}
\label{results}

\subsection{$\chi^2$ analysis}
\label{chi2}

We determine the values of 14 parameters
from the best fit to the $A_1(x, Q^2)$ data
for the proton ($p$), neutron ($n$) and deuteron ($d$).
Using the GRV parametrization for the unpolarized PDFs at the LO and
NLO \cite{GRV98} and the SLAC measurement of $R(x, Q^2)$,
we construct $A_1^{\rm calc}(x,Q^2)$ for the $p$, $n$, and $d$.
For the deuteron, we use
$g_1^d = \frac{1}{2} (g_1^p + g_1^n )(1-\frac{3}{2} \omega_D)$
with the D-state probability in the deuteron $\omega_D = 0.05$.

Then, the best parametrization is obtained by minimizing
$\chi^2=\sum(A_1^{\rm data}(x,Q^2)-A_1^{\rm calc}(x,Q^2) )^2/
(\Delta A_1^{\rm data}(x,Q^2) )^2$ with
{\sc Minuit} \cite{minuit}, where $\Delta A_1^{\rm data}$
represents the error on the experimental data including both
systematic and statistical errors.
Since some of the systematic errors are correlated, it leads
to an overestimation of errors to include all systematic
errors. On the other hand, if we fully exclude them, the
uncertainties in the experimental data are not properly
reflected in the analysis. Because of our choice to include
the systematic errors, the $\chi^2$ defined in our analysis is not
properly normalized. The minimum $\chi^2$ divided by a number
of degree-of-freedom achieved in the analysis is often smaller
than unity. Consequently the $\chi^2$ in our analysis should
be regarded as only a relative measure of the fit to
the experimental data. In addition, the parameter errors
are overly estimated.
We have confirmed that inclusion of only statistical errors in
the $\chi^2$ analysis does not change the results significantly
except a change of the $\chi^2$ by 7\%, which is
consistent with the change of the error size.

In evolving the distribution functions with $Q^2$,
we neglect the charm-quark contributions
to $A_1(x,Q^2)$ and take the flavor number $N_f=3$
because the $Q^2$ values of the $A_1$ experimental data are not
so large compared with the charm threshold. To be consistent with
the unpolarized, we use the same values as the GRV,
$\Lambda_{\rm QCD}^{(3)}=\ 204 \ {\rm MeV}$ at LO and
$\Lambda_{\rm QCD}^{(3)}=\ 299 \ {\rm MeV}$ at NLO in
the $\overline {\rm MS}$ scheme.
The NLO scale parameter leads to the value of $\alpha_s(M_Z^2)=0.118.$ 
In order to obtain a solution which satisfies the positivity condition,
we make further refinements to the parametrization functions $h_i(x)$.
The technical details are discussed in Appendix \ref{appen-pos}.

\begin{figure}[h]
      \vspace{1.0cm}
   \begin{center}
       \epsfig{file=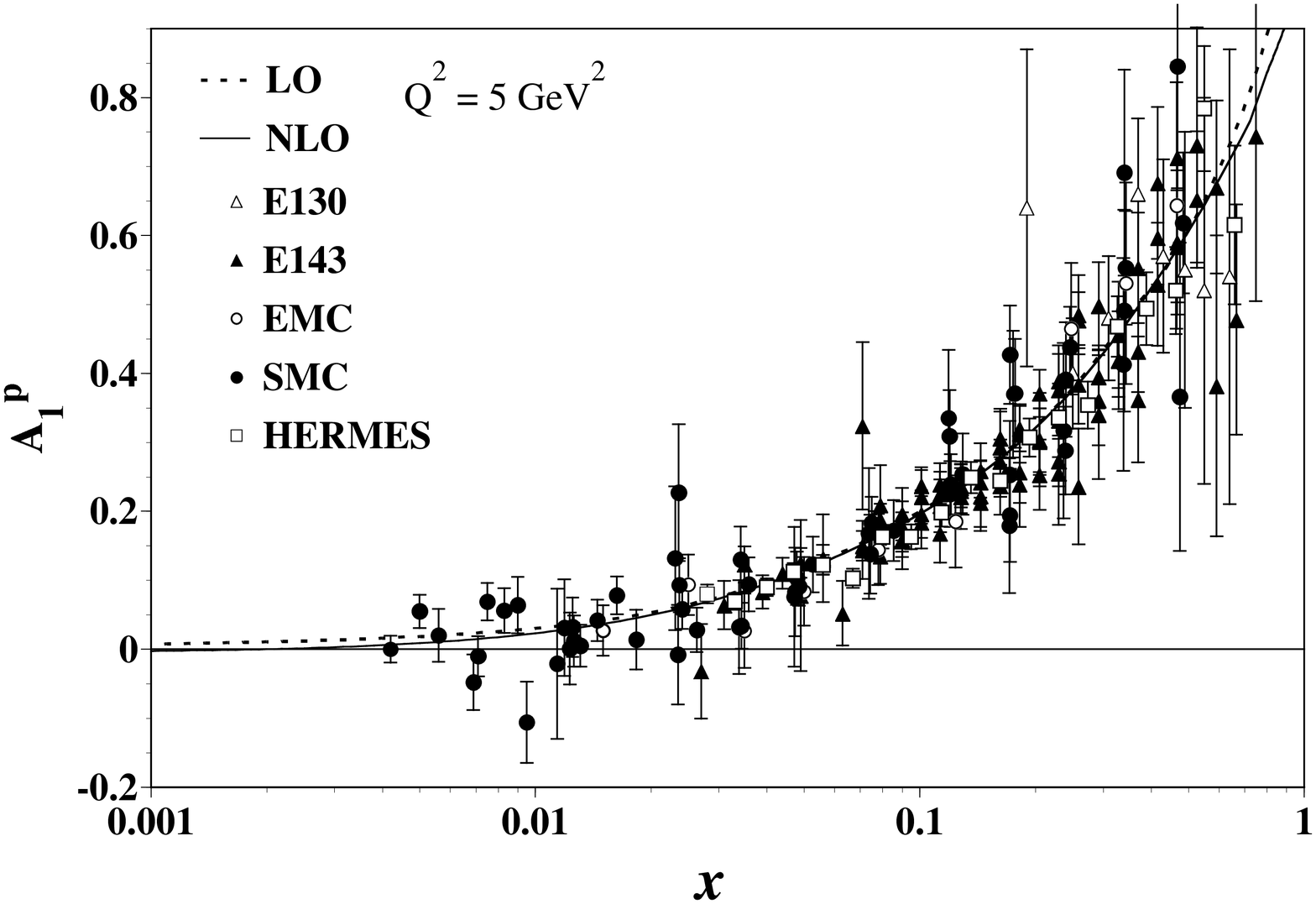,width=14.0cm} \\
           (a)
   \end{center}
   \vspace{+1.5cm}
   \begin{center}
       \epsfig{file=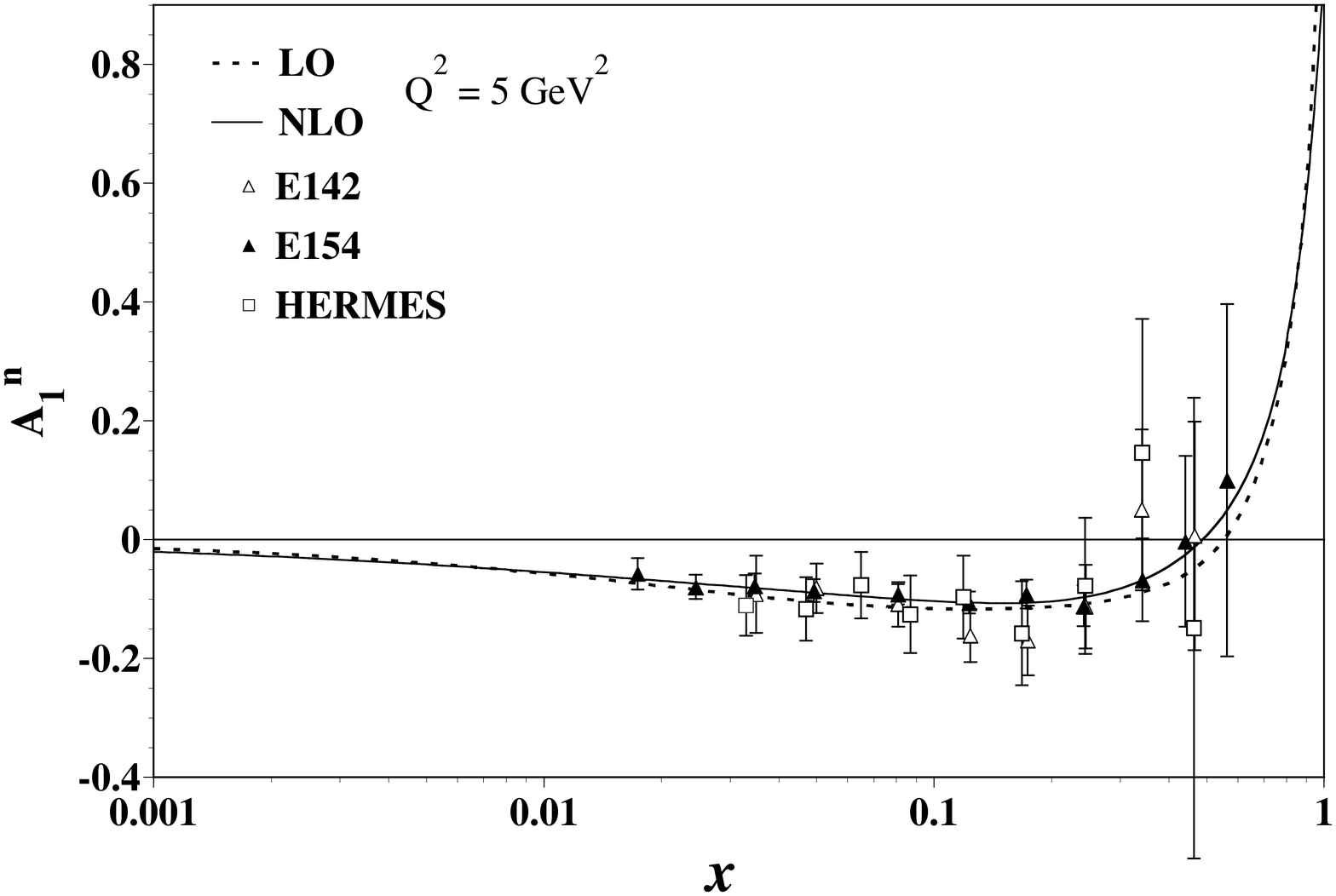,width=14.0cm} \\
           (b)
   \end{center}
   \vspace{0.0cm}
\vfill\eject
   \begin{center}
       \epsfig{file=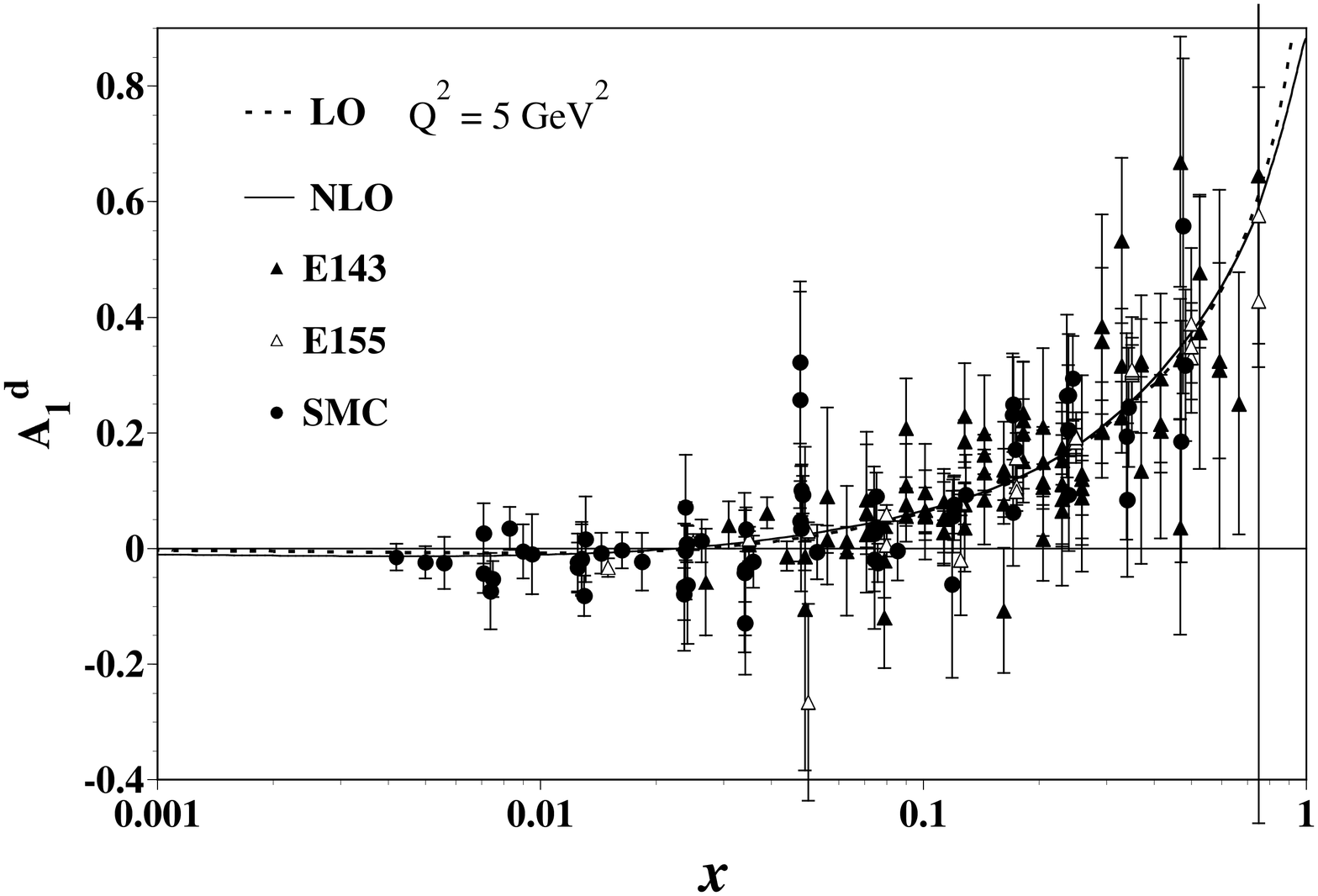,width=14.0cm} \\
           (c)
   \end{center}
   \vspace{1.0cm}
   \caption{Comparison of our calculations with the experimental
            asymmetry $A_1(x, Q^2)$ data for the (a) proton, (b) neutron,
            and (c) deuteron.
            Our results are obtained at $Q^2$=5 GeV$^2$
            with the optimum parameters in Tables \ref{T:LO} (LO)
            and \ref{T:NLO} (NLO-1).
            The NLO and LO results are shown
            by the solid and dotted lines, respectively.}
   \label{fig:a1}
\end{figure}

The results are presented in
Table \ref{T:LO} for the LO with $\chi^2$/d.o.f=322.6/360
and in Table \ref{T:NLO} for the NLO
with $\chi^2$/d.o.f=300.4/360.
Because the values of $A_i$ are determined by the first moments
for the $\Delta u_v$ and $\Delta d_v$ distributions, they are listed
without errors. 
We show the LO and NLO fitting results for the asymmetry $A_1$
together with experimental data in Fig. \ref{fig:a1}.
The theoretical curves are calculated at $Q^2$=5 GeV$^2$.
The asymmetries are shown for the (a) proton, (b) neutron, 
and (c) deuteron. 
As the experimental data, the E130, E143, EMC, SMC, and HERMES
proton data are shown in Fig. \ref{fig:a1}(a); 
the E142, E154, and HERMES neutron data are in (b);
the E143, E155, and SMC deuteron data are in (c). Kinematical conditions
and analysis methods of these experiments are listed
in Table \ref{T:Exp}.
We find from these figures that the obtained parameters reproduce
well the experimental data of $A_1$ in both LO and NLO cases.
However, there are slight differences between the LO and NLO curves
in Fig. \ref{fig:a1}, and three factors contribute to the differences.
First, the most important difference is the contribution
of the polarized gluon distribution through the coefficient 
function. Second, the LO and NLO evolutions are different
because not only the the splitting functions
but also the scale parameters are different.
Third, the LO and NLO expressions are different
in the unpolarized GRV distributions.


\subsection{Comparison of LO and NLO analyses}
\label{lo-nlo}

Comparing the value of $\chi^2$/d.o.f for the LO with that for the
NLO, we found a better description of the experimental data with
the NLO analysis. The value of $\chi^2$/d.o.f is improved
by 7\%. This implies that it is necessary to analyze
the data in the NLO if one wants to get better information
on the spin structure of the nucleon from the polarized DIS data.

\begin{figure}[h]
      \vspace{1.0cm}
   \begin{center}
       \epsfig{file=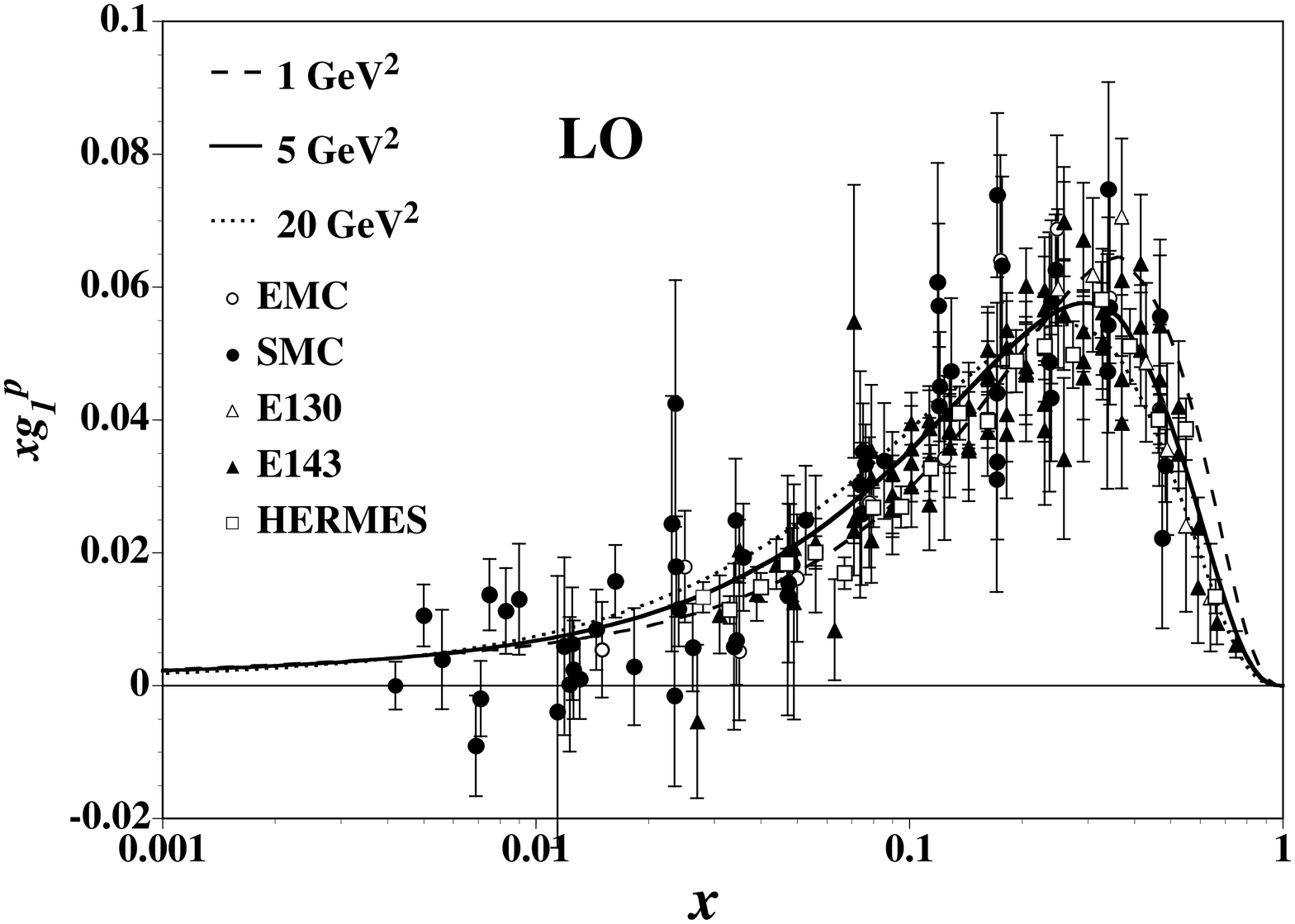,width=14.0cm} \\
           (a)
   \end{center}
   \vspace{1.5cm}
   \begin{center}
       \epsfig{file=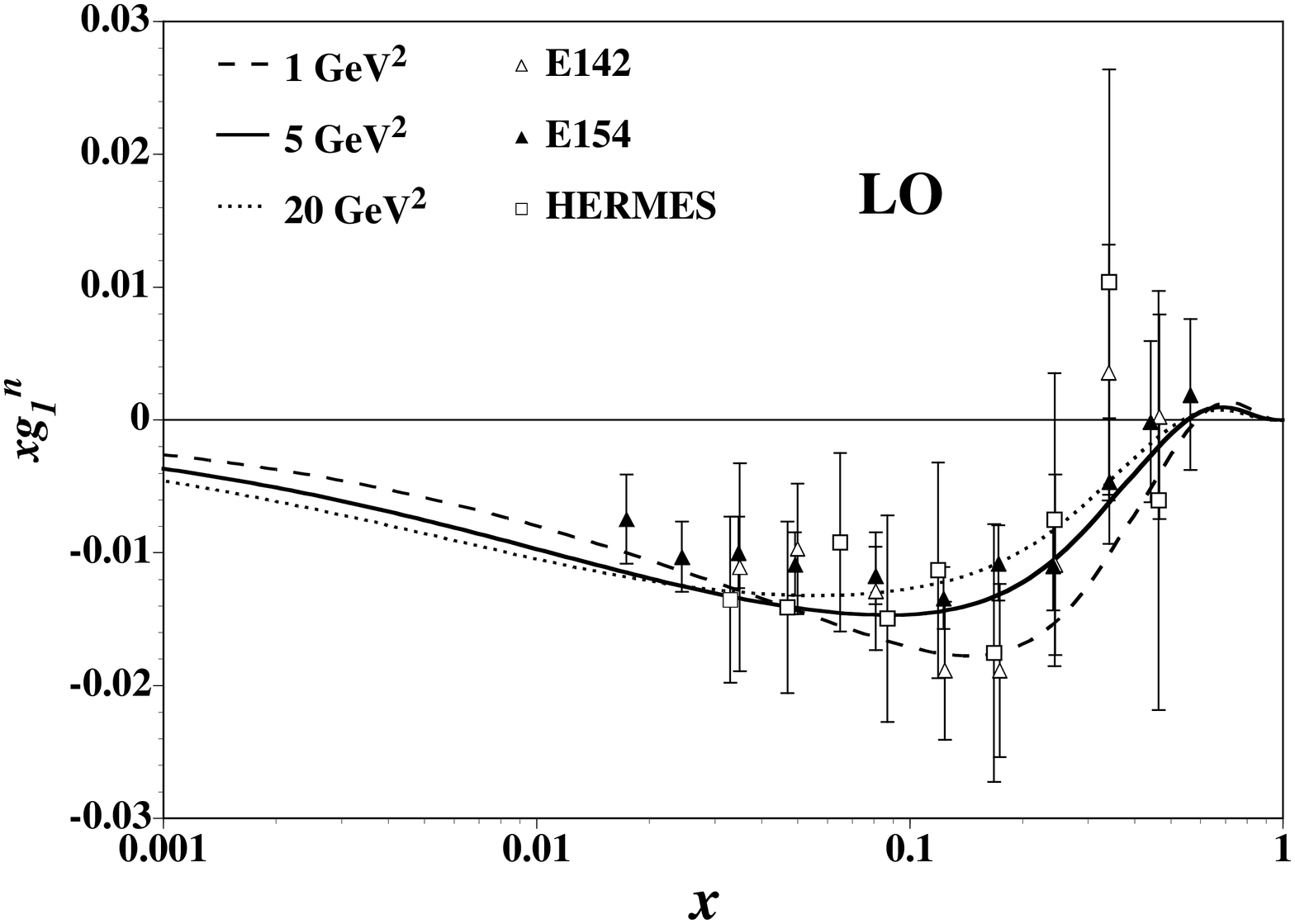,width=14.0cm} \\
           (b)
   \end{center}
   \vspace{0.0cm}
\vfill\eject
   \begin{center}
       \epsfig{file=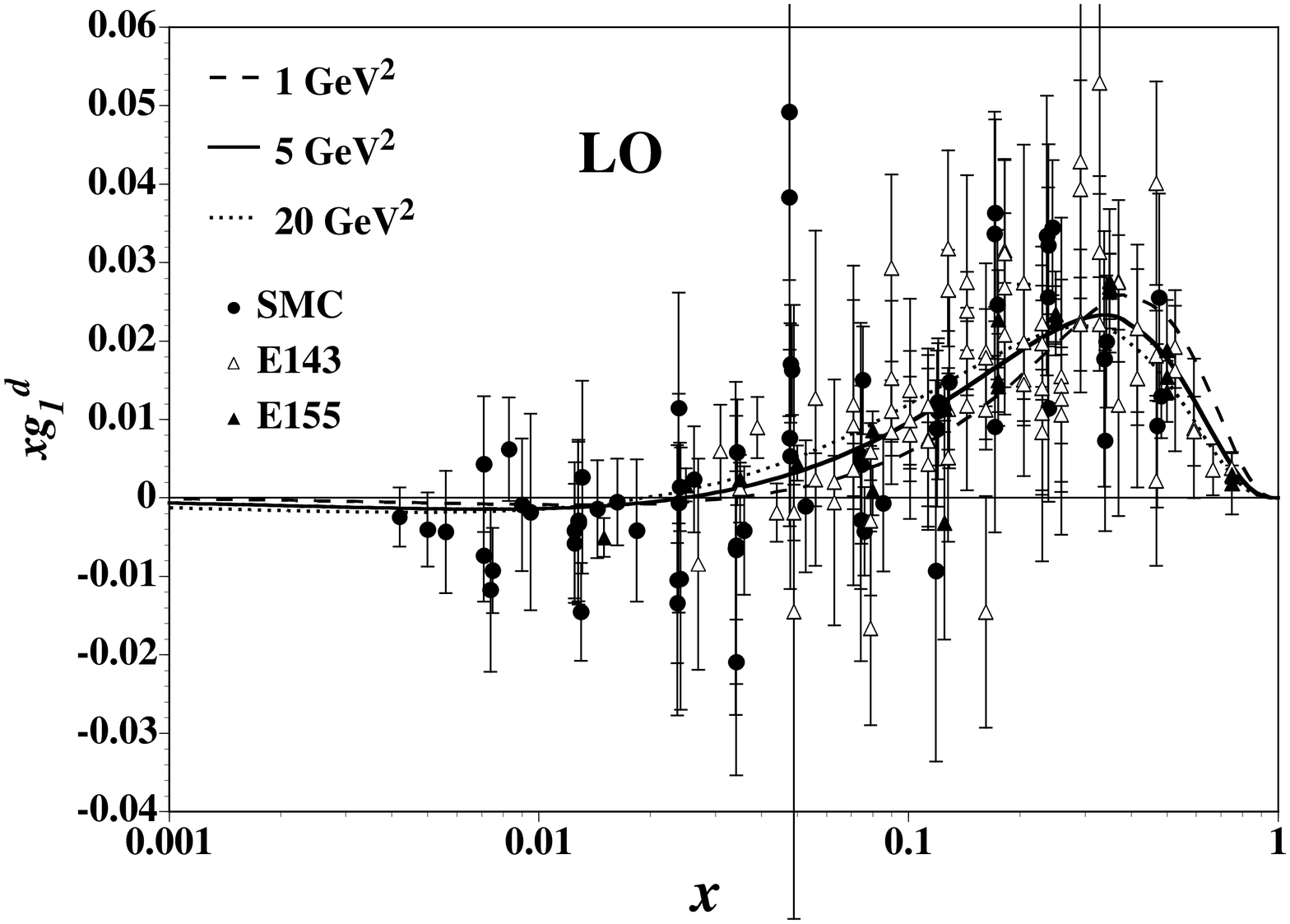,width=14.0cm} \\
           (c)
   \end{center}
   \vspace{1.0cm}
   \caption{Experimental spin-dependent structure functions $xg_1(x, Q^2)$
            are compared with our LO results for the (a) proton,
            (b) neutron, and (c) deuteron. Our fitting results are
            calculated at $Q^2$=1, 5, 20 GeV$^2$ by using the LO evolution
            equations with the optimum parameters in Table \ref{T:LO},
            and they are shown by the dashed, solid, and dotted
            curves, respectively. The experimental data are obtained
            from the $A_1(x, Q^2)$ data and
            the $F_2(x, Q^2)$ calculated with the unpolarized GRV
            distributions and $R_{1990}(x, Q^2)$.}
   \label{fig:g1lo}
\end{figure}

\begin{figure}[h]
      \vspace{1.0cm}
   \begin{center}
       \epsfig{file=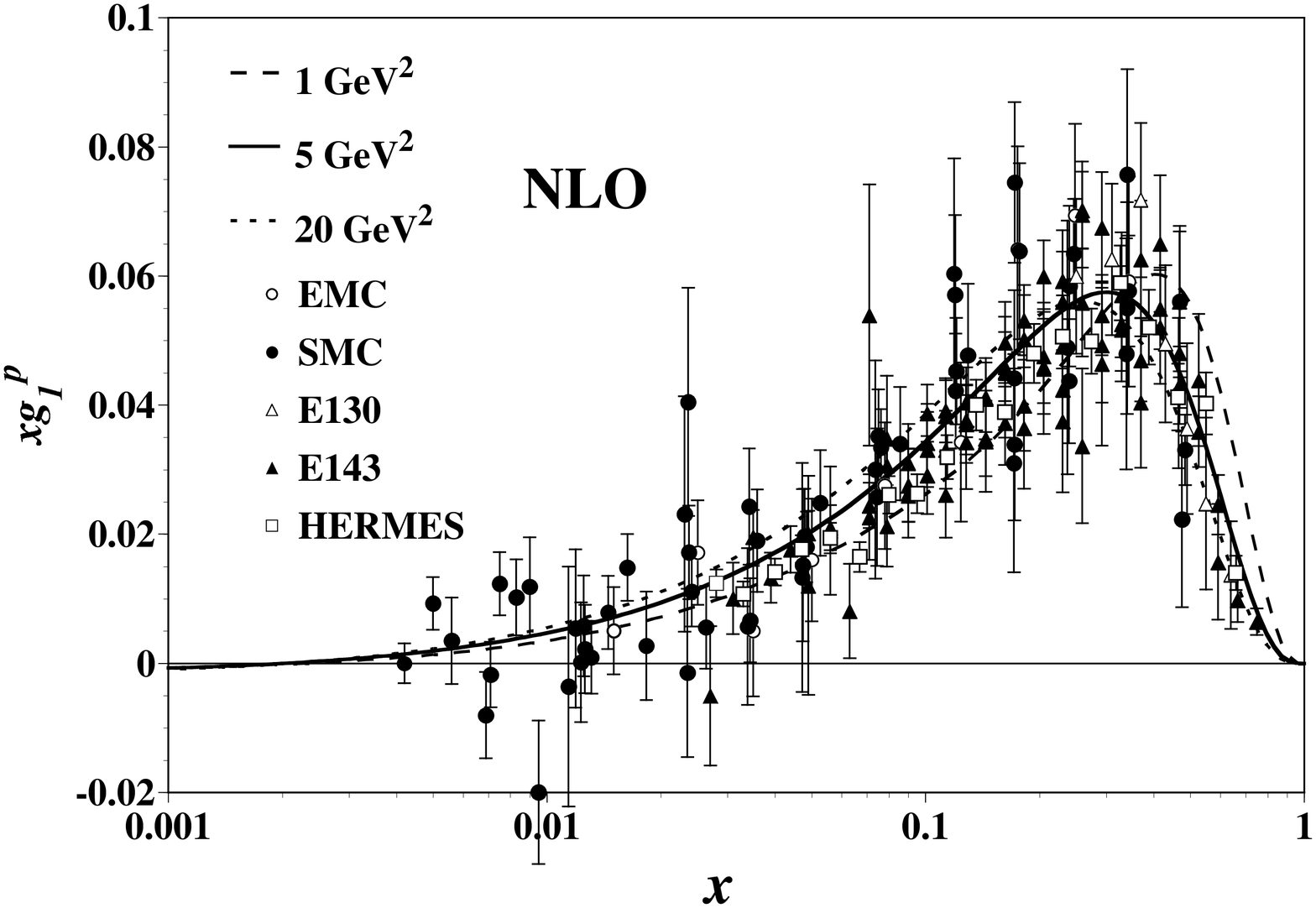,width=14.0cm} \\
           (a)
   \end{center}
   \vspace{1.5cm}
   \begin{center}
       \epsfig{file=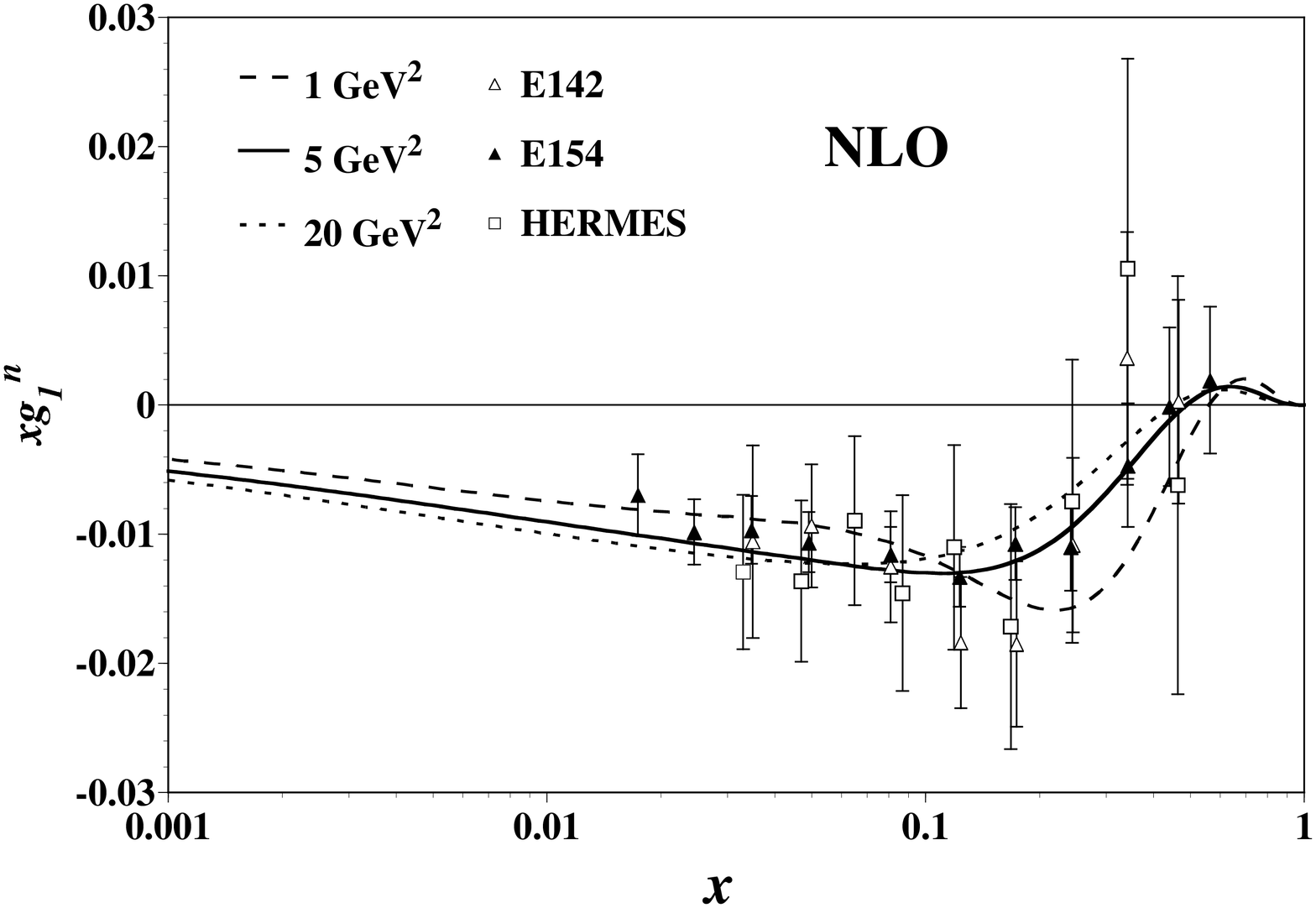,width=14.0cm} \\
           (b)
   \end{center}
   \vspace{0.0cm}
\vfill\eject
   \begin{center}
       \epsfig{file=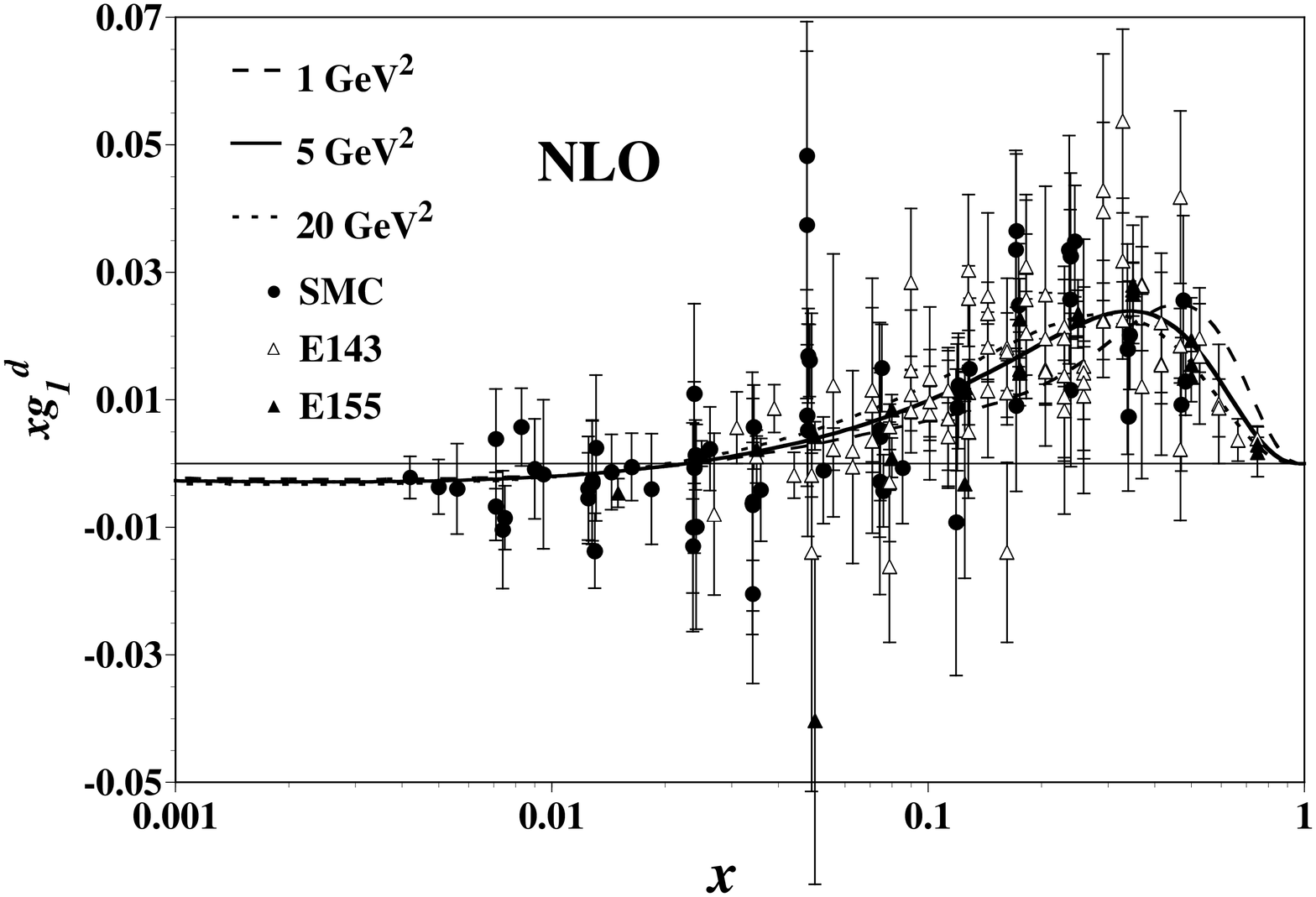,width=14.0cm} \\
           (c)
   \end{center}
   \vspace{1.0cm}
   \caption{Experimental data of $xg_1(x, Q^2)$
            are compared with our NLO-1 results for the (a) proton,
            (b) neutron, and (c) deuteron.
            The notations are the same as those in Fig. \ref{fig:g1lo}.}
   \label{fig:g1nlo}
\end{figure}

The $\chi^2$ contribution from each data set is listed
in Table \ref{T:chi2}. The improvement is significant especially
for the HERMES proton and E154 neutron data.
The results of $g_1$ at the LO and NLO are shown in Fig. \ref{fig:g1lo}
and Fig. \ref{fig:g1nlo}, respectively.
The ``experimental" $g_1$ data are calculated by using 
Eqs. (\ref{eqn:a1}) and (\ref{eqn:f1}) together with the raw data
for the asymmetry $A_1$ and the GRV unpolarized distributions.
The theoretical results are shown by the dashed, solid, and dotted
curves at $Q^2$=1, 5, 20 GeV$^2$. As already shown in Fig. \ref{fig:xg1},
the $g_1$ structure function shifts to the smaller-$x$ region as
$Q^2$ increases. It is rather difficult to discuss the agreement
with the deuteron data in Figs. \ref{fig:g1lo}(c) and \ref{fig:g1nlo}(c)
because of the large experimental errors. However, the proton and
neutron data at small $x$ tend to agree with the theoretical
curves at $Q^2$=1 GeV$^2$. It is particularly clear in the neutron
$g_1$ in Figs. \ref{fig:g1lo}(b) and \ref{fig:g1nlo}(b).
Furthermore, the proton, neutron, and deuteron data at large $x$
agree with the LO and NLO curves at $Q^2$=20 GeV$^2$. 
There are correspondences of the data to the theoretical results
because the small-$x$ data are typically in the small-$Q^2$
range ($Q^2=1 \sim$a few GeV$^2$) and the large-$x$ data are
in the large-$Q^2$ range ($Q^2 \gtrsim $10 GeV$^2$). 

As seen in Figs. \ref{fig:g1lo} and \ref{fig:g1nlo},
the LO $g_1^{\, p}$ is slightly larger at small $x$ in comparison
with the NLO $g_1^{\, p}$, while the LO $g_1^{\, n}$ is
smaller than the NLO $g_1^{\, n}$ in the range $0.01<x<0.2$.
The NLO fit agrees better with the data.
The $\chi^2$ improvement in the NLO for the HERMES and E154
data in Table \ref{T:chi2} is explained as follows by using
Fig. \ref{fig:a1}.
In comparing the theoretical curves with the data, we should note
that the theoretical asymmetries are given at fixed $Q^2$
($Q^2$=5 GeV$^2$), whereas the data are at various $Q^2$ values.
However, as it is found in Fig. \ref{fig:a1}(a),
the LO curve is slightly above the NLO one and also the HERMES data. 
It makes the $\chi^2$ value larger in the LO analysis.
In Fig. \ref{fig:a1}(b), it is clear that the LO curve
deviates from the E154 neutron data, so that the $\chi^2$ contribution
becomes larger from the E154 data. 
It is well known that the difference between the NLO
($\overline{\rm MS}$ scheme) and LO originates from
the polarized gluon contribution to the structure function $g_1$
via the Wilson coefficient.
Accordingly, the result that the NLO fit is better than the LO
implies that the polarized gluon has a nonzero contribution to the nucleon
spin, $i.e.$ $\Delta g\ne 0$ at $Q^2_0$.
Furthermore, we find in this analysis that the NLO fit 
is more sensitive to the polarized gluon distribution than the LO one.
Therefore, we can conclude that the NLO analysis is necessary to extract
information on the polarized gluon distribution.


\subsection{Behavior of polarized parton distribution functions}

\begin{figure}[h]
      \vspace{0.0cm}
   \begin{center}
       \epsfig{file=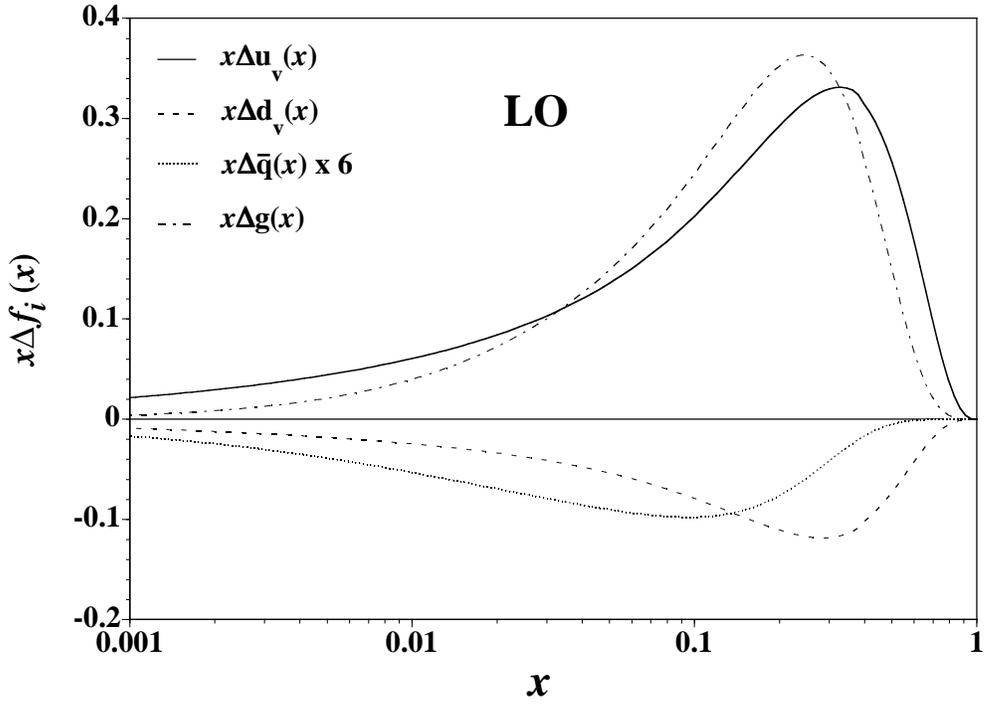,width=14.0cm} \\
           (a)
   \end{center}
   \vspace{0.3cm}
   \begin{center}
       \epsfig{file=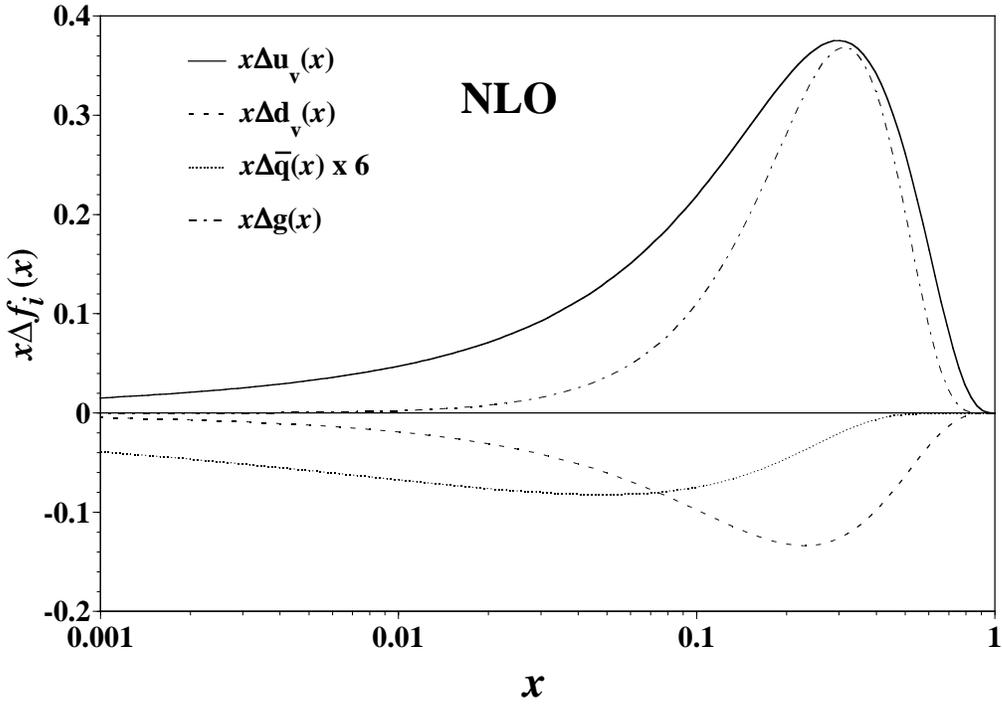,width=14.0cm} \\
           (b)
   \end{center}
   \vspace{0.3cm}
   \caption{Obtained LO and NLO-1 polarized parton distributions
            $x\Delta f_i(x, Q^2)$ 
            at $Q^2=1$ GeV$^2$ in (a) and (b), respectively.}
   \label{fig:df}
\end{figure}

We show the behavior of polarized parton distributions
$x\Delta f_i(x, Q^2)$ as a function of $x$ at $Q^2=1$ GeV$^2$
for the (a) LO and (b) NLO cases in Fig. \ref{fig:df}.
The first moment for $\Delta u_v (x)$ is fixed at the positive value
($\eta_{u_v}$=0.926) and the one for $\Delta d_v (x)$
is at the negative value ($\eta_{d_v}=-$0.341), so that
the obtained distributions $\Delta u_v(x)$ and $\Delta d_v(x)$
become positive and negative, respectively.
In the same way as the other $\chi^2$-analysis results,
the antiquark (gluon) distribution becomes negative (positive)
at small- and medium-$x$ regions. The gluon distribution cannot be
determined well by only the lepton scattering data. 
In particular, the gluon distribution plays a role 
in $g_1$ only through the $Q^2$ evolution in the LO, so that
$\Delta g(x)$ cannot be uniquely determined. Even if it is neglected
in the analysis ($\Delta g=0$), the $\chi^2$ difference is not so
significant in the LO. The NLO effects are apparent by comparing
Fig. \ref{fig:df}(a) with Fig. \ref{fig:df}(b).
In the NLO, the gluon distribution contributes to $g_1$ additionally
through the coefficient function; therefore, it modifies the
valence-quark distributions (particularly the $\Delta u_v$) and
the antiquark distribution. The NLO distribution $\Delta u_v$
becomes significantly smaller than the LO one at small $x$,
and the NLO distribution $\Delta \bar q$ becomes a more singular
function as $x\rightarrow 0$. Because of more involvement
of the gluon distribution in $g_1$, the determination of $\Delta g$
is better in the NLO $\chi^2$ analysis. 

\begin{figure}[h]
      \vspace{0.0cm}
   \begin{center}
       \epsfig{file=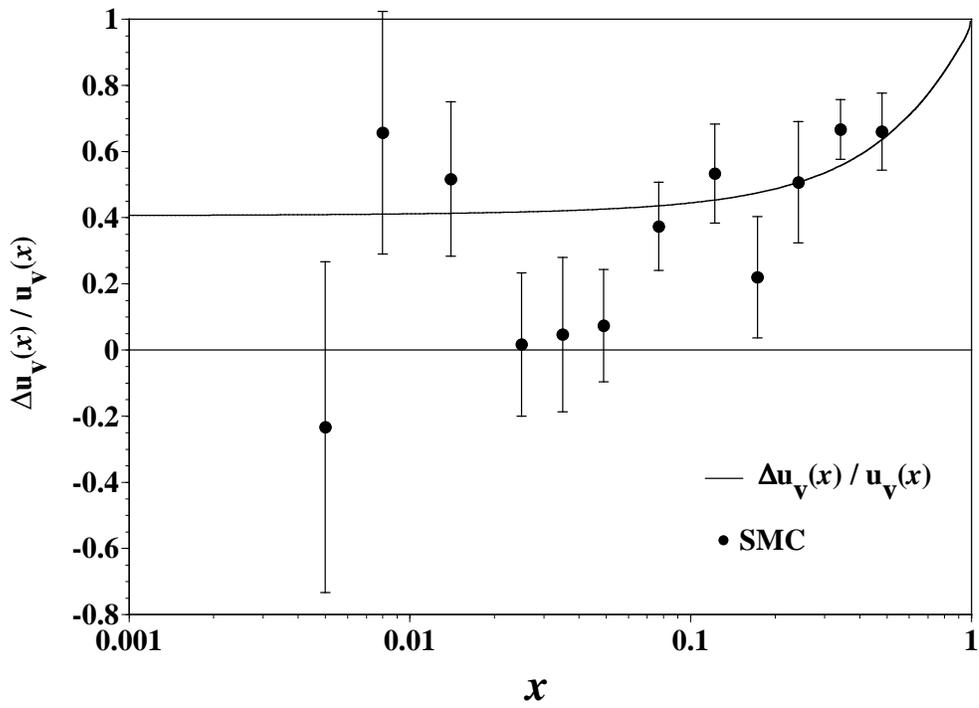,width=14.0cm} \\
           (a)
   \end{center}
   \vspace{0.3cm}
   \begin{center}
       \epsfig{file=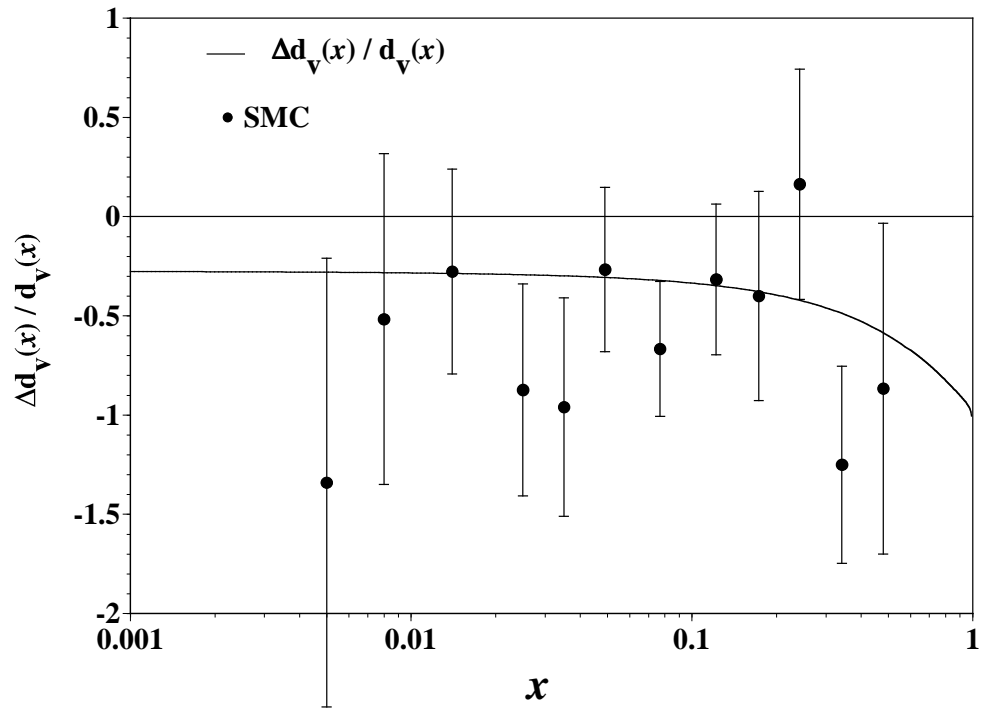,width=14.0cm} \\
           (b)
   \end{center}
   \vspace{0.3cm}
   \caption{Our LO ratios $\Delta u_v/u_v$ and $\Delta d_v/d_v$
            are compared with the SMC data in (a) and (b), respectively.
            Our results are calculated at $Q^2=10$ GeV$^2$ by using
            the polarized distributions in the LO.}
   \label{fig:dfsmc}
\end{figure}

Recently, the measurement of polarized parton distributions of each
flavor has been carried out by the SMC
in semi-inclusive processes of the polarized DIS \cite{SMC-incl}.
Although we did not include the semi-inclusive date in our analysis
from the consideration of the data precision and the analysis framework,
it is still possible to compare our polarized PDFs with their analysis.  
In order to compare with the SMC data,
the LO initial distributions are evolved to those at $Q^2=10$ GeV$^2$
by the LO evolution equations. Then, the ratios $\Delta u_v(x)/u_v(x)$
and $\Delta d_v(x)/d_v(x)$ are shown in Fig. \ref{fig:dfsmc}
together with the SMC data. The theoretical ratios are roughly
constants in the small-$x$ region ($x<0.1$) and 
$\Delta u_v(x)/u_v(x)$ approaches $+1$ as $x\rightarrow 1$ whereas
$\Delta d_v(x)/d_v(x)$ approaches $-1$.
We find that our LO parametrization
seems to be consistent with the data. 
However, it is unfortunate that our NLO parametrization cannot
be compared with the data since the SMC data are analyzed only
for the LO.


\subsection{Small-$x$ behavior of polarized antiquark distributions}
\label{smallx}

As we obtained in the $\chi^2$ analyses, the small-$x$ behavior
of the parton distributions is controlled by the parameter $\alpha$.
It is obvious from Tables \ref{T:LO} and \ref{T:NLO} that
the small-$x$ behavior cannot be determined in the antiquark
and gluon distributions.
For example, the obtained parameter is listed as
$\alpha_{\bar q} (NLO) = 0.32 \pm 0.22$ with a large error.
It suggests that the small-$x$ part of the antiquark distribution
cannot be fixed by the existing data. In order to clarify
the situation, we need to have higher-energy facilities such as
polarized-HERA and eRHIC \cite{erhic}. 

Because the present experimental data are not enough for determining
the small-$x$ behavior, we should consider to fix the parameter $\alpha$
for the antiquark distribution by theoretical ideas.
The gluon parameter $\alpha_g$ cannot be also determined. However, we
leave the problem for future studies because the lepton scattering
data are not sufficient for determining the gluon distribution in any case.
Some predications are made for $\alpha_{\bar q}$ in the following by using 
the Regge theory and the perturbative QCD. 

According to the Regge model, the structure function $g_1$ 
in the small-$x$ limit is controlled by the intercepts ($\alpha$) of
$a_1 (1260)$, $f_1 (1285)$, and $f_1 (1420)$ trajectories:
\begin{equation}
g_1 (x)  \sim  x^{-\alpha} \ \ \ \ {\rm as} \ x\rightarrow 0.
\end{equation}
However, not only the $a_1$ intercept but also the $f_1$ intercepts are
not well known. It is usually assumed as $\alpha_{a_1}=-0.5 \sim 0$
\cite{regge}. Therefore, we expect
$\Delta \bar q  \sim  x^{(0.0, \, 0.5)}$,
where $x^{(0.0, \, 0.5)}$ indicates that the function is
in the range from $x^{0.0}$ to $x^{0.5}$.
Since our parametrization is provided for
the function $h_i(x)=\Delta f_i(x)/f_i(x)$,
we should find out the small-$x$ behavior
of the unpolarized distribution. According to our numerical analysis,
the GRV distribution has the property $x \, \bar q \sim x^{-0.14}$
at $Q^2$=1 GeV$^2$. Taking these small-$x$ functions into account,
the Regge prediction is
\begin{equation}
h_{\bar q}^{Regge} (x) \sim x^{(1.1, \, 1.6)} ,
\label{eqn:regge-x}
\end{equation}
if the theory is applied at $Q^2=1$ GeV$^2$.
Our LO and NLO fits result in $x^{0.59}$ and $x^{0.32}$,
respectively, as $x\rightarrow 0$.
These functions look very different from
Eq. (\ref{eqn:regge-x}); however, they are not inconsistent
if the errors of Tables \ref{T:LO} and \ref{T:NLO} are
taken into account.  

The perturbative QCD could also suggest the small-$x$ behavior.
In the small-$x$ limit, the splitting functions are dominated
by the most singular terms. Therefore, if we can assume that
the singlet-quark and gluon distributions are constants at certain
$Q^2$ ($\equiv Q_1^2$) in the limit $x\rightarrow 0$,
their singular behavior is predicted from the evolution equations.
According to its results, the singlet distribution behaves like
\cite{lr-sum}
\begin{equation}
\Delta \Sigma (x,Q^2) \sim \exp \left[ 
        2 \sqrt{ \frac{8 \, C_A}{\beta_0} \,
                   \xi (Q^2) \, \ln \frac{1}{x} } \, \right] ,
\label{eqn:pqcd-1}
\end{equation}
where $\xi(Q^2)=\ln [\alpha_s(Q_1^2)/\alpha_s(Q^2)]$, $C_A=3$,
and $\beta_0=11-2N_f/3$. The problem is to find an
appropriate $Q_1^2$ where the singlet and gluon distributions
are flat at small $x$. Choosing the range $Q_1^2=0.3 \, \sim \, 0.5$
GeV$^2$ and $Q^2=1$ GeV$^2$, we fit the above equation numerically
by the functional form of $x^{-\alpha}$ at small $x$. Then, the obtained
function is in the range, $x^{(-0.12, \, -0.09)}$.
Because the unpolarized distribution is given by
$x \, \bar q \sim x^{-0.14}$, the perturbative QCD
(with the assumption of the above $Q_1^2$ range) suggests
\begin{equation}
h_{\bar q}^{pQCD} (x) \sim x^{1.0} .
\label{eqn:pqcd-2}
\end{equation}
This function falls off much faster than ours at small $x$.

In this way, we found that the perturbative QCD and the Regge theory
suggest the small-$x$ distribution as
$h_{\bar q} \sim x^{(1.0,1.6)}$.
Because the small $x$ behavior cannot be determined by the $\chi^2$ analyses
in Sec. \ref{chi2}, we had better fix the power of $x$ by these theoretical
implications. In this subsection, the NLO $\chi^2$ analyses are reported by
fixing the parameter at $\alpha_{\bar q}$=0.5, 1.0, and 1.6.
The middle value is the perturbative QCD estimate, and the latter two
ones are roughly in the Regge prediction range. The first one is taken
simply by considering a slightly singular distribution than these
theoretical predictions.

The obtained parameters and $\chi^2$ are listed in Table \ref{T:NLOfix}.
Considering the NLO value $\chi^2$=300.4 in Table \ref{T:chi2},
we find that the $\chi^2$ change is 0.1\%, 1.8\%, and 7.7\% for 
$\alpha_{\bar q}$=0.5, 1.0, and 1.6, respectively.
The $\chi^2$ changes are so small in $\alpha_{\bar q}$=0.5
and 1.0 that they could be equally taken as good parametrizations
in our studies. Using the obtained distributions with fixed
$\alpha_{\bar q}$, we have the first moments and spin contents in
Table \ref{T:1stmom-fix}. Because of the small-$x$ falloff
for larger $\alpha_{\bar q}$, the antiquark first moment and spin
content change significantly. If the perturbative QCD and Regge
prediction range ($\alpha_{\bar q}$=1.0 and 1.6) is taken, the calculated
spin content is within the usually quoted values $\Delta\Sigma=0.1\sim 0.3$.
The obtained $\chi^2$ value suggests that the $\alpha_{\bar q}$=1.0
solution could be also taken as one of the good fits to the data.
In this sense, our results are not inconsistent with
the previous analyses. However, the results indicate that a better
solution could be obtained for smaller $\alpha_{\bar q}$, so that the spin
content could be smaller than the usual values
$\Delta\Sigma=0.1\sim 0.3$. At least, we can state that
the present data are not taken at small enough $x$,
so that the spin content cannot be determined uniquely.

We found that the $\alpha_{\bar q}$=0.5 and 1.0 results could be 
also considered as good parametrizations to the experimental data.
The $\chi^2$ is so large in the $\alpha_{\bar q}$=1.6 analysis 
that its set cannot be considered a good fit to the data.
Because the $\alpha_{\bar q}$=0.5 results are almost the same as
the NLO ones in Sec. \ref{lo-nlo}, it is redundant to take it
as one of our parametrizations. Therefore, we propose
the LO and NLO distributions (sets: LO and NLO-1) in \ref{lo-nlo}
together with the $\alpha_{\bar q}$=1.0 distributions (set: NLO-2)
as three sets of the AAC parametrizations. 

Although the parametrization for $\Delta f_i / f_i$ is necessary
for imposing the positivity condition, it is rather cumbersome
for practical applications in calculating other cross sections
in the sense that we always need both our parametrization results
and the GRV unpolarized distributions at $Q^2$=1 GeV$^2$. 
Furthermore, it is not convenient that the analytical GRV
distributions are not given at $Q^2$=1 GeV$^2$.
In Appendix \ref{appendix}, we supply simple functions for
the three AAC distributions without resorting to the GRV
parametrization for the practical calculations.


\subsection{Spin contents of polarized quarks and gluons}

The first moment of each polarized parton distribution
and the integrated $g_1$ at $Q^2$=1, 5, and 10 GeV$^2$ are given
in Table \ref{T:1stmom} for the LO and NLO.
At $Q^2=1$ GeV$^2$, the amounts of quarks and gluons
carrying the nucleon spin are
\begin{eqnarray}
\Delta\Sigma &=& 0.201, \ \ \ \ \Delta g=0.831,
\ \ \ \ \text{in the LO}, 
\nonumber \\
\Delta\Sigma &=& 0.051, \ \ \ \ \Delta g=0.532,
\ \ \ \ \text{in the NLO-1}, \\
\Delta\Sigma &=& 0.241, \ \ \ \ \Delta g=0.533,
\ \ \ \ \text{in the NLO-2}. 
\nonumber
\end{eqnarray}
These results confirm that the quarks carry a small amount of
the nucleon spin.
The first moments of the structure functions at $Q^2=1$ GeV$^2$ are
\begin{eqnarray}
\Gamma^p_1(Q^2) &=& 0.144, \ \ \ \ \Gamma^n_1(Q^2)=-0.067,
\ \ \ \ \Gamma^d_1(Q^2)=0.036,
\ \ \ \ \text{in the LO}, 
\nonumber \\
\Gamma^p_1(Q^2) &=& 0.110, \ \ \ \ \Gamma^n_1(Q^2)=-0.069,
\ \ \ \ \Gamma^d_1(Q^2)=0.019,
\ \ \ \ \text{in the NLO-1}, \\
\Gamma^p_1(Q^2) &=& 0.128, \ \ \ \ \Gamma^n_1(Q^2)=-0.051,
\ \ \ \ \Gamma^d_1(Q^2)=0.035,
\ \ \ \ \text{in the NLO-2}.
\nonumber
\end{eqnarray}
Because the first moment of $\Delta u_v - \Delta d_v$
is fixed by Eq. (\ref{eqn:UVDV}), the Bjorken sum rule is
satisfied in both LO and NLO at any $Q^2$
within the perturbative QCD range.

\begin{figure}[h]
      \vspace{0.0cm}
   \begin{center}
       \epsfig{file=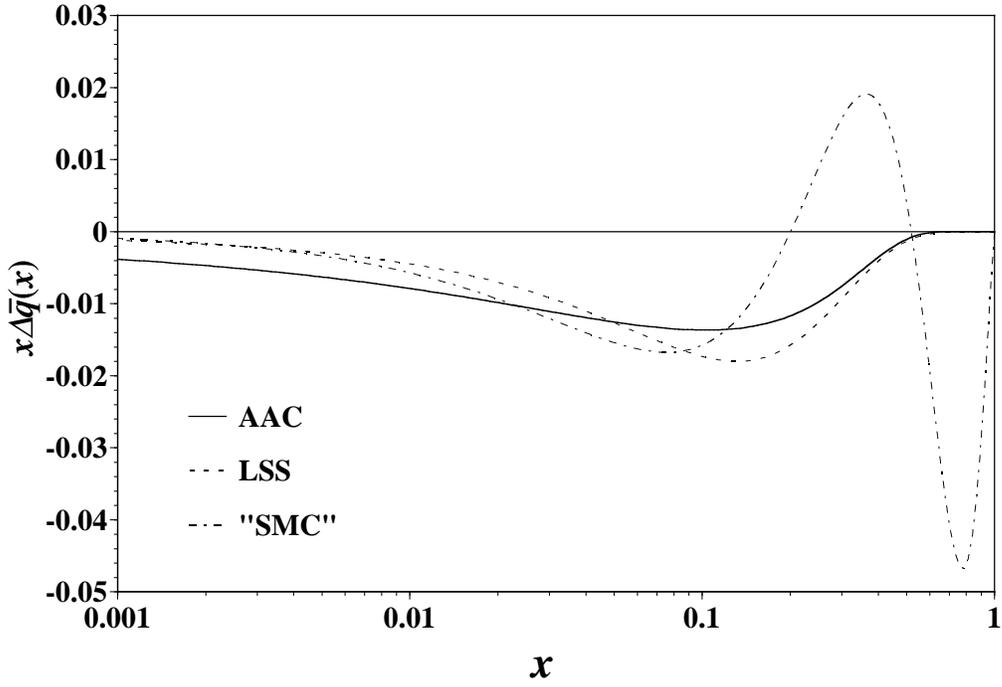,width=14.0cm}
   \end{center}
   \vspace{-0.2cm}
 \caption{The antiquark distributions of transformed SMC (``SMC") and
          LSS(1999) are compared with our NLO-1 distribution
          at $Q^2$=1 GeV$^2$.}
 \label{fig:dqbar}
\end{figure}

It should be noted that our $\Delta \Sigma$ in the NLO-1
seems to be considerably smaller than the usual values
published so far in many other papers. In fact, the recent SMC and
Leader-Sidrov-Stamenov (LSS) parametrizations
\cite{SMCfit,LSS} obtained $\Delta\Sigma=$0.19 and 0.28, respectively,
at $Q^2$=1 GeV$^2$.
The difference originates mainly from the small-$x$ behavior of
the antiquark distribution. We compared our NLO-1 distribution $\Delta\bar q$,
which is denoted as AAC, with the other $\overline{\rm MS}$ distributions 
in Fig. \ref{fig:dqbar}. The LSS(1999) antiquark distribution is directly
given in their parametrization, whereas the SMC distribution is calculated
by using their singlet and nonsinglet distributions.
Because the antiquark distribution is not directly given in the SMC
analysis, we may call it as a transformed SMC (``SMC") distribution.
The transformed SMC has peculiar $x$ dependence at medium and
large $x$; however, all the distributions
agree in principle in the region ($0.01<x<0.1$) where accurate experimental
data exist and the antiquark distribution plays an important role.
On the other hand, it is clear that our distribution does not fall off
rapidly as $x \rightarrow 0$ in comparison with the others.
This is the reason why our NLO-1 spin content is significantly smaller. 

\begin{figure}[h]
      \vspace{0.0cm}
   \begin{center}
       \epsfig{file=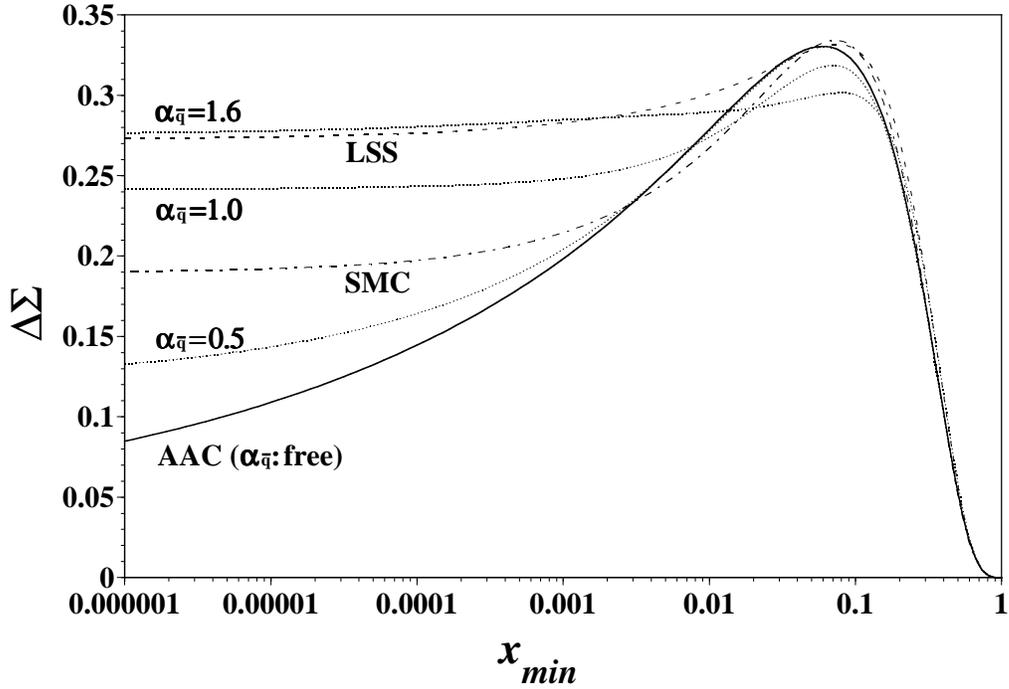,width=14.0cm}
   \end{center}
   \vspace{-0.2cm}
   \caption{The $x_{min}$ dependence of
            $\Delta \Sigma (x_{min})=\int _{x_{min}}^1 \Delta \Sigma (x)dx$
            is compared with the recent parametrizations
            of SMC and LSS(1999) at $Q^2$=1 GeV$^2$.}
   \label{fig:dsigma}
\end{figure}

In order to clarify the difference, we plot the spin content
in the region between $x_{min}$ and 1 by calculating
$\Delta \Sigma (x_{min})=\int _{x_{min}}^1 \Delta \Sigma (x)dx$
in Fig. \ref{fig:dsigma}. Because the LSS and SMC distributions
are less singular functions of $x$, their spin contents saturate
even at $x=10^{-4}$ although our $\Delta \Sigma$ still decreases
in this region. The difference simply
reflects the fact that the accurate experimental data are not
available at small $x$. The parametrization results with
fixed $\alpha_{\bar q}$ are also shown. As the antiquark distribution
becomes less singular, the spin content becomes larger. 
As mentioned in the previous subsection, the $\alpha_{\bar q}$=1.0
results could be taken as a good fit. The spin content is 0.24
in this case and it is completely within the usual range
$\Delta\Sigma=0.1 \sim 0.3$. 

The small-$x$ issue has been discussed in other publications.
The idea itself stems from the publication of Close and Roberts
\cite{cr}, and it is also noted in the numerical analyses of
Altarelli, Ball, Forte, and Ridolfi (ABFR) \cite{ABFR}.
In the ABFR parametrization, various fits are tried by assuming
the small-$x$ behavior, and they obtain the first moment of
$a_0(x)$ as $a_0=0.02 \sim 0.18$.
Therefore, our NLO-1 analysis is consistent with their studies
although the spin content seems to be smaller than the
usual one ($0.1 \sim 0.3$). In this way, our NLO-1 analysis
result may seem very different from many other publications,
it is essentially consistent with them. 
It indicates that the small-$x$ ($\sim 10^{-5}$) data are
absolutely necessary for the determination of the spin content.


\subsection{Comparison with recent parametrizations}

We have already partially discussed the comparison with the
recent parametrization results in the previous subsection.
However, the detailed discussions are necessary particularly
on the differences between these analyses in order 
to clarify the difference in the physical basis.

First, we discuss differences between our parametrization and the LSS.
Before the detailed comparison,
we used their $\chi^2$-fitting procedure in our program and
confirmed their numerical results. It indicates that both
fitting programs are consistent although evolution methods and
other subroutines are completely different.   

Our parametrization functions are similar to theirs.
In fact, both methods use the parametrization for the ratio
of the polarized distribution to the unpolarized one
($\Delta f_i (x)/f_i (x)=h_i (x)$, $i=u_v, \ d_v, \ \bar q, \ g$). 
The LSS parametrization employed a very simple function
$h_i(x)=A_i x^{\alpha_i}$, and we used a more complicated one
$h_i(x)=A_i x^{\alpha_i} (1+\gamma_i x^{\lambda_i})$.
This may seem to be insignificant; however, the extra parameters
provide wide room for the functions to readjust in the $\chi^2$
analysis. According to our studies, the minimum $\chi^2$ cannot
reach anywhere close to our minimum point if the LSS
function is used in our fit. 
Therefore, although it is a slight modification,
the outcome has a significant difference.
Furthermore, the LSS gluon distribution fails to satisfy
the positivity condition at large $x$ although it does not
matter practically at this stage.

Another important difference is how to calculate the spin
asymmetry $A_1$ from the unpolarized distributions. 
There are two issues in this calculation procedure.
One is that LSS kept the factor $1+4 M_N^2 x^2/Q^2$
in handling the SLAC data, 
whereas we neglected. Another is that LSS calculated 
the structure function $F_1$ directly from the unpolarized
distributions, whereas we calculated it by Eq. (\ref{eqn:f1}).
As for the first point, we have checked that inclusion
of the factor has not significant impact on the results.
It is partly because the factor
$1+4 M_N^2 x^2/Q^2$ modifies the asymmetry $A_1$ at large $x$
but the $Q^2$ values are generally large in such a $x$ region.
The second point is more serious. Their method is right
in the light of perturbative QCD. However, the $F_2$ structure
functions are generally used rather than $F_1$ in obtaining
the unpolarized PDFs. If there were no higher-twist
contributions, it does not matter whether $F_1$
is calculated directly or Eq. (\ref{eqn:f1}) is used.
However, it is well known that the higher-twist effects
are rather large as obvious from the function $R(x,Q^2)$
in the SLAC-1990 analysis \cite{R1990}. 
It modifies the asymmetries as large as 35\%, and the modification
is conspicuous in the whole $x$ region. In the LSS analysis,
perturbative QCD contributions to the function $R$ are included
due to the coefficient-function difference between $F_1$ and $F_2$,
but they are small in the small- and medium-$x$ regions. 
This difference in handling $F_1$ creates the discrepancy between
the LSS and our polarized antiquark distributions, and it is
especially important for determining their small-$x$ behavior.

Next, we discuss comparison with the SMC parametrization.
Our $\chi^2$ analysis is different from theirs in the parametrization
functions. We parametrized the ratios $\Delta f_i /f_i$
($i=u_v$, $d_v$, $\bar q$, $g$).
As mentioned in Sec. \ref{parton}, the analysis by the SMC
in Ref. \cite{SMCfit} utilized the separation of the polarized
quark distributions into ($\Delta \Sigma(x)$, $\Delta q_{\rm NS}^{p}(x)$,
and $\Delta q_{\rm NS}^{n}(x)$) which can, in principle, be
transformed into $\Delta u^+ (x)$, $\Delta d^+ (x)$, and $\Delta s^+ (x)$.

When we do this transformation of SMC results to compare
with the polarized sea-quark distributions from our analysis and LSS,
we find that the polarized strange-quark distribution ($\Delta s(x)$)
from the ``transformed SMC" oscillates as shown in Fig. \ref{fig:dqbar}.
However, this simply implies that the conventionally
used functional form has a limitation and the distribution
functions obtained from different separations can be
quite different. The uncertainty of the sea-quark distribution
was also pointed out in the analysis by Gordon, Goshtasbpour,
and Ramsey \cite{GGR}. We should re-emphasize that
direct measurement of the sea-quark polarization is
very important. At the highest energy of polarized $pp$
collisions at RHIC, the weak bosons are copiously produced
and the parity violating asymmetry ${\cal A}_{L}$ for its production
is very useful in elucidating spin-flavor structure of the nucleon
\cite{SAITO}. With such direct measurement, the uncertainty
in the polarized sea-quark distribution will be much reduced.

Common differences from the SMC and LSS are that a large set of
data tables is used for $A_1$ rather than the $Q^2$ averaged one.
Although the present data may not have the accuracy to discuss the
$Q^2$ dependence, it is desirable to use the large table if one
wishes to obtain better information on the gluon distribution.
Furthermore, an advantage of our results is that the positivity
condition is strictly satisfied, so that our parametrizations
does not pose any serious problem in practical applications.


\section{CONCLUSIONS}
\label{concl}

We have analyzed the experimental data for the spin asymmetry $A_1$
of the proton, neutron, and deuteron by using a simple parametrization
for the ratios of polarized parton distributions to the corresponding
unpolarized ones. We discussed the details on physical meanings behind
our parametrization and also on our $Q^2$ evolution method. 
As a consequence, we found that the asymmetry $A_1$ could have significant
$Q^2$ dependence in the small $Q^2$ region ($Q^2<2$ GeV$^2$), so that
frequently-used assumption of the $Q^2$ independence in $A_1$ cannot be
justified in a precise analysis. From the LO and NLO $\chi^2$ analyses,
we obtained good fits to the experimental data. Because the NLO $\chi^2$
is significantly smaller than that of LO, the NLO analysis should be
necessarily used in the parametrization studies. An advantage of our
analysis is that the positivity condition is satisfied in the whole $x$
region. An important consequence of our analyses is that the small-$x$
behavior of the sea-quark distributions cannot be uniquely determined by
the present data, so that the usual spin content 
$\Delta \Sigma =0.1 \sim 0.3$ could be significantly modified
depending on the future experimental data at small $x$ ($\sim 10^{-5}$).
Our LO and NLO analyses suggested $\Delta\Sigma$=0.20 and 0.05, respectively.
However, if we take theoretical suggestions by ``perturbative QCD"
and Regge theory for the polarized antiquark distribution at small $x$, 
the spin content becomes $\Delta \Sigma =0.24 \sim 0.28$ in the NLO.
The obtained gluon distributions are positive in both LO and NLO,
but it is particularly difficult to determine $\Delta g$ in the LO.
From these analyses, we have proposed one LO set and two NLO sets
of parametrizations as the AAC polarized parton distribution
functions.


\section*{{\bf Acknowledgments}}
\addcontentsline{toc}{section}{\protect\numberline{\S}{ACKNOWLEDGMENTS}}

The authors would like to thank A. Br\"ull, M. Grosse-Perdekamp, V. Hughes,
R. L. Jaffe, K. Kobayakawa, and D. B. Stamenov for useful discussions
or email communications.
This work has been done partly within the framework of RIKEN RHIC-Spin
project, and it was partly supported by the Japan Society for the Promotion
of Science and also by the Japanese Ministry of Education, Science,
and Culture.



\appendix

\section{Treatment of positivity condition \\
          in our $\chi^2$ analysis}
\label{appen-pos}

Additional modification of the function $h_i(x)$
is desirable in the actual $\chi^2$ fitting.
Although Eq. (\ref{eqn:gx2}) is a useful functional form,
it is not very convenient for the $\chi^2$ analysis in the sense that
the positivity condition is rather difficult to be satisfied.
In fact, running our $\chi^2$ program, we obtain a solution which
does not necessarily meet the positivity requirement.
In order to take into account this condition, the function is 
slightly modified although it is equivalent in principle:
\begin{eqnarray}
 \begin{aligned}
  h_i(x) &= \xi_i x^{\nu_i} + \kappa_i x^{\mu_i} \\
         &= \delta_i x^{\nu_i}-\kappa_i (x^{\nu_i} - x^{\mu_i}),
         \ \ \ \ \ i=u_v, \, d_v, \, \bar q, \, g
\label{eqn:gx3}
 \end{aligned}
\end{eqnarray}
where $\delta_i=\xi_i + \kappa_i$.
It can be seen why this function is more suitable at $x=1$ by
the following simple example.
The original function is given by two parameters,
$h_i (x=1)=A_i \, (1+\gamma_i)$; however,
the modified one is by only one parameter $h_i (x=1)=\delta_i$.
Therefore, it is more easier to restrict the function $h_i (x)$ within
the positivity-condition range.
There is another advantage that the parameters are rather
independent each other. For example, the parameter $\lambda_i$ is
strongly correlated with $\alpha_i$ ($\lambda_i \ge - \alpha_i$)
if we would like to avoid singular behavior as $x\rightarrow 0$.
In this way, the functional form of Eq. (\ref{eqn:gx3}) is used
in the actual $\chi^2$ fitting although it is mathematically 
equivalent to Eq. (\ref{eqn:gx2}).

Although we could perform the $\chi^2$ analysis with the supplied
information, it is not straight forward to obtain a solution
which satisfies the positivity condition. We describe the details
of the analysis procedure. First, it was already mentioned that
the first moments of $\Delta u_v$ and $\Delta d_v$ are fixed by the $F$
and $D$ values, and they are given by
\begin{equation}
  \eta_i = \int_0^1 dx \, [\delta_i x^{\nu_i}
                   -\kappa_i (x^{\nu_i} - x^{\mu_i})] \,  f_i(x) 
  \ \ \ \ (i=u_v ,\ d_v) .
\label{eqn:1stmom}
\end{equation}
Then, the parameters $\kappa_{u_v}$ and $\kappa_{d_v}$ are
determined by
\begin{equation}
  \kappa_i = \frac{\delta_i\int dx \, x^{\nu_i} f_i(x) -\eta_i}
                  {\int dx \, (x^{\nu_i}-x^{\mu_i}) f_i(x)} .
\end{equation}
As we explained in Sec. \ref{smallx}, theory
suggests the functions $h_i$ should not be a singular function of
$x$ in the small-$x$ region. Therefore, we try to find a solution
in the parameter range $\mu_i, \nu_i \ge 0$.

Next, we discuss the positivity condition.
If the signs of the parameters $\xi_i$ and $\kappa_i$ are
the same, the function $h_i(x)$ is a monotonically increasing
or decreasing function, so that $h_i(x=1)=\delta_i$ should be within
the range $-1\le \delta_i \le +1$ due to the positivity requirement.
On the other hand, if the signs are different, the function could have
an extreme value at certain $x$ ($\equiv X$). If $X$ is larger than one,
the function could be a monotonic one in the range ($0\le x\le 1$).
Then, the same condition $-1\le \delta_i \le +1$ is applied.
However, if $X$ is smaller than
one, the situation is slightly complicated. Because the first and
second terms have the same functional form in the first equation of
Eq. (\ref{eqn:gx3}), we can have either $\mu_i<\nu_i$ or $\mu_i>\nu_i$.
Therefore, the condition $\mu_i<\nu_i$ is taken (practically
only for $\Delta \bar q$ and $\Delta g$)
in the following analysis without loosing generality.
From Eq. (\ref{eqn:gx3}), we find that the extreme value is located at 
\begin{equation}
X = \left( -\frac{\kappa_i \zeta_i}{\xi_i} 
       \right)^{\frac{1}{\nu_i-\mu_i}},
\end{equation}
where $\zeta_i=\mu_i/\nu_i$ ($0<\zeta_i<1$).
It is in the range $0<X<1$ if the condition
$0 < - \kappa_i \zeta_i /\xi_i <1$, namely
\begin{equation}
   \begin{array}{rl}
   \frac{\delta_i}{1-\zeta_i} < \kappa_i& \ \ \mbox{for $0<\kappa_i$} , \\
   \frac{\delta_i}{1-\zeta_i} > \kappa_i& \ \ \mbox{for $\kappa_i<0$} ,
   \end{array}
\label{eqn:kappamin}
\end{equation}
is satisfied. The extreme value is then obtained as 
\begin{equation}
h_i(X) = \left( - \frac{\kappa_i \, \zeta_i}{\delta_i - \kappa_i}
         \right) ^\expb \, \kappa_i(1-\zeta_i) .
\label{eqn:extreme}
\end{equation}
Using the positivity condition $|h_i(X)| \le 1$, we obtain
the following constraint on the parameters:
\begin{equation}
g^{+}(\kappa_i) \equiv \kappa_i-\delta_i -\kappa_i \zeta_i 
                  \left[ \kappa_i(1-\zeta_i) \right]^\expbi \ge 0 ,
\label{eqn:con+1}
\end{equation}
in the case $\kappa_i >0$ ($0<h_i(X) \le 1$).
Because the function $g^{+}(\kappa_i)$ has a positive curvature,
we try to find a $\kappa_i$ point ($=\kappa_i'$) which satisfies
$g^{+}(\kappa_i ')=0$. There is only one solution
for negative $\delta_i$ and two solutions for positive $\delta_i$.
In any case, we seek the solution $\kappa_i '$ which is larger than
the extreme point $\kappa_i=1/(1-\zeta_i)$ by the Newton's method.
Then, the parameter $\kappa_i$ is redefined as
$\kappa_i=\sigma_i \kappa_i '$. The parameters $\sigma_i$ are
used in the $\chi^2$ analysis for the antiquark and gluon distributions
within the range $0 \le \sigma_i \le 1$, so that the actual functional
form is
\begin{equation}
h_i(x) = \delta_i x^{\alpha_i}-{\sigma_i \kappa_i^{\prime}} 
(x^{\alpha_i} - x^{\alpha_i \zeta_i}) 
\ \ \ \ {\rm for} \ i=\bar q, \ g.
\label{eqn:hsea}
\end{equation}
On the other hand, we find 
\begin{equation}
g^{-}(\kappa_i) \equiv \kappa_i-\delta_i -\kappa_i \zeta_i 
                  \left[ - \kappa_i(1-\zeta_i) \right]^\expbi \le 0 ,
\label{eqn:con-1}
\end{equation}
in the case $\kappa_i <0$ ($-1 \le h_i(X) < 0$).
A similar analysis is done for the function $g^{-}(\kappa_i)$
in order to satisfy the positivity condition.
With these preparations, we can perform the $\chi^2$ analysis.

\vfill\eject


\section{Practical polarized parton distributions}
\label{appendix}

Our polarized parton distributions are given in the parametrized
functions $h_i(x)$ multiplied by the GRV unpolarized distributions.
For practical applications, we supply the following three sets of
simple functions, which reproduce the $\chi^2$ analysis results in Sec.
\ref{results}, as the AAC distributions at $Q^2$=1 GeV$^2$:
\begin{align}
{\bf \rm Set: AAC-LO}  &
\nonumber \\
x\Delta u_v (x) & = 0.4949 \, x^{0.456} (1-x)^{2.84} (1+9.60 \, x^{1.23}),
\ \ \ \ \ \ \ \ \ \ \ \ \
\nonumber \\
x\Delta d_v (x) & = -0.2040 \, x^{0.456} (1-x)^{3.77} (1+14.6 \, x^{1.36}),
\nonumber \\
x\Delta \bar q (x) & = -0.1146 \, x^{0.536} (1-x)^{10.5}
                                            (1+39.4 \, x^{1.93}),
\nonumber \\
x\Delta g (x) & = 2.738 \, x^{0.908} (1-x)^{5.61} (1+12.3 \, x^{1.60}) ,
\end{align}
\begin{align}
{\bf \rm Set: AAC-NLO-1}  &
\nonumber \\
x\Delta u_v (x) & = 0.4029 \, x^{0.478} (1-x)^{3.18} (1+15.1 \, x^{1.07}),
\ \ \ \ \ \ \ \ \ \ \ \ \
\nonumber \\
x\Delta d_v (x) & = -0.2221 \, x^{0.568} (1-x)^{3.92} (1+9.46 \, x^{0.813}),
\nonumber \\
x\Delta \bar q (x) & = -0.03249 \, x^{0.230} (1-x)^{7.77}
                                         (1+3.65 \, x^{0.883}),
\nonumber \\
x\Delta g (x) & = 8.844 \, x^{1.77} (1-x)^{6.21} (1+13.6 \, x^{1.51}) ,
\end{align}
\begin{align}
{\bf \rm Set: AAC-NLO-2}  &
\nonumber \\
x\Delta u_v (x) & =0.4353 \, x^{0.465} (1-x)^{2.94} (1+8.98 \, x^{0.938}),
\ \ \ \ \ \ \ \ \ \ \ \ \
\nonumber \\
x\Delta d_v (x) & = -0.1850 \, x^{0.471} (1-x)^{3.89} (1+14.0 \, x^{1.11}),
\nonumber \\
x\Delta \bar q (x) & = -0.2452 \, x^{0.752} (1-x)^{8.13} ,
\nonumber \\
x\Delta g (x) & = 8.895 \, x^{1.77} (1-x)^{6.22} (1+13.6 \, x^{1.51}) .
\end{align}

\vfill\eject


\vfill\eject


\begin{table}
\addcontentsline{toc}{section}{\protect\numberline{\S}{TABLES}}
\caption{Summary of published polarized DIS experimental data on
         the spin asymmetry $A_1$. The listed are the number of
         data points above $Q^2$=1 GeV$^2$.}
\label{T:Exp}
\begin{center}
\begin{tabular}{l|cccccc}
Exp.  &  $x$ range  & $Q^2$ range & \#of data  & $A_2/g_2(x,Q^2)$
&$R(x,Q^2)$ & Ref. \\
      &              &              & 
      & contribution
&     & \\
\hline
EMC  ($p$) & 0.015$-$0.466 & 3.5$-$29.5~GeV$^2$ &10 & neglected & $Q^2$-indep &
\cite{EMC}\\
SMC  ($p$) & 0.005$-$0.480 & 0.25$-$72.07~GeV$^2$ & 12 & neglected & $R_{\rm
1990}$ & \cite{SMC98}\\
E130 ($p$) & 0.18$-$0.70 & 3.5$-$10.0~GeV$^2$  & 8
           & neglected & constant &
\cite{E130}\\
E143 ($p$) & 0.022$-$0.847 & 0.28$-$9.53~GeV$^2$ & 81
           & measured & $R_{\rm 1990}$ & \cite{E14398}\\
HERMES ($p$) & 0.021$-$0.85 & 0.8$-$20.0~GeV$^2$ & 19 &
      E143/SMC & $R_{\rm 1990}$ & \cite{HERMES}\\
\hline
SMC  ($d$) & 0.005$-$0.480 & 1.3$-$54.4~GeV$^2$ & 12
           & neglected & $R_{\rm 1990}$
& \cite{SMC98} \\
E143 ($d$) & 0.022$-$0.847 & 0.28$-$9.53~GeV$^2$ & 81
           & measured & $R_{\rm 1990}$
& \cite{E14398}\\
E155 ($d$) & 0.01$-$0.9 & 1.0$-$40.0~GeV$^2$ & 24 &
        $g_2^{WW}$  & $R_{\rm 1990}$/NMC & \cite{E155}\\
\hline
E142 ($n$) & 0.035$-$0.466 & 1.1$-$5.5~GeV$^2$ & 8 
           & neglected & $R_{\rm 1990}$ &
\cite{E142} \\
E154 ($n$)  & 0.0174$-$0.5643 & 1.21$-$15.0~GeV$^2$ & 11
            & measured & $R_{\rm
1990}$ & \cite{E154}\\
HERMES ($n$) \ & 0.033$-$0.464 & 1.22$-$5.25~GeV$^2$ & 9 
               & neglected & $R_{\rm
1990}$ & \cite{HERMES}\\
\end{tabular}
\end{center}
\end{table}

\vspace{-0.5cm}
\begin{table}
\caption{Obtained parameters at $Q^2=1$ GeV$^2$
          in the leading-order $\chi^2$ analysis.}
\label{T:LO}
\begin{center}
\begin{tabular}{c|cccc}
distribution \  & $A$                        & $\alpha$
                & $\gamma$                   & $\lambda$  \\
\hline
$\Delta u_v$    &     0.404 $\pm$   0.054  &   0.00 $\pm$   0.01
                &     1.47  $\pm$   0.20   &   1.41 $\pm$   0.46 \\
$\Delta d_v$    &  $-$0.274 $\pm$   0.056  &   0.00 $\pm$   0.01
                &     2.65  $\pm$   0.54   &   1.25 $\pm$   0.28 \\
$\Delta \bar q$ &  $-$0.680 $\pm$   0.373  &   0.59 $\pm$   0.94
                &  $-$2.47  $\pm$   0.82   &   4.06 $\pm$   1.14 \\
$\Delta g$      &    47.5   $\pm$   4.1    &   1.44 $\pm$   0.73
                &  $-$0.986 $\pm$   0.002  &   0.06 $\pm$   1.05 \\
\end{tabular}
\end{center}
\end{table}

\vspace{-0.5cm}
\begin{table}
\caption{Obtained parameters at $Q^2=1$ GeV$^2$
          in the next-to-leading-order $\chi^2$ analysis (set NLO-1).}
\label{T:NLO}
\begin{center}
\begin{tabular}{c|cccc}
distribution \  & $A$                        & $\alpha$
                & $\gamma$                   & $\lambda$  \\
\hline
$\Delta u_v$    &     0.356 $\pm$   0.047  &   0.00  $\pm$   0.00
                &     1.54  $\pm$   0.20   &   0.889 $\pm$   0.058 \\
$\Delta d_v$    &  $-$0.502 $\pm$   0.031  &   0.153 $\pm$   0.065
                &     0.992 $\pm$   0.062  &   2.48  $\pm$   0.27  \\
$\Delta \bar q$ &  $-$0.269 $\pm$   0.107  &   0.32  $\pm$   0.22
                &  $-$4.72  $\pm$   1.48   &   3.20  $\pm$   0.47  \\
$\Delta g$      &   249.2   $\pm$   8.3    &   2.15  $\pm$   0.11
                &  $-$1.0040 $\pm$  0.0002 &   0.031 $\pm$   0.152 \\
\end{tabular}
\end{center}
\end{table}

\vspace{-0.5cm}
\begin{table}[hbt]
\caption{ \protect{$\chi^2$} contribution of experimental data compared
          with the number of data points.  Here, the NLO indicates
          the set NLO-1 .}
\label{T:chi2}
\begin{center}
\begin{tabular}{c|ccc}
experimental \ & \# of data & \multicolumn{2}{c}{$\chi^2$} \\
 data \ &  ($Q^2 >$1.0~GeV$^2$) &  LO &  NLO  \\
\hline
EMC ($p$)      & 10 & 5.2  & 4.6    \\
SMC ($p$)      & 59 & 55.0 & 53.7   \\
E130 ($p$)     & 8  & 5.1  & 5.2     \\
E143 ($p$)     & 81 & 65.0 & 60.8  \\
HERMES ($p$) \ & 19 & 23.1 & 17.2  \\
\hline
SMC ($d$)      & 65 & 56.6 & 54.0  \\
E143 ($d$)     & 81 & 79.1 & 81.2  \\
E155 ($d$)     & 24 & 20.0 & 17.1  \\
\hline
E142 ($n$)     & 8  & 3.5  & 2.4    \\
E154 ($n$)     & 11 & 7.5  & 1.8    \\
HERMES ($n$)   & 9  & 2.6  & 2.3    \\
\hline
total          & 375 & 322.6 & 300.4 \\
\end{tabular}
\end{center}
\end{table}

\vspace{-0.5cm}
\begin{table}
\caption{Obtained parameters at $Q^2=1$ GeV$^2$
         in the NLO $\chi^2$ analysis with fixed $\alpha_{\bar q}$.}
\label{T:NLOfix}
\begin{center}
\begin{tabular}{c|ccccc}
distribution \   & $\chi^2$ 
                 & $A$                        & $\alpha$
                 & $\gamma$                   & $\lambda$  \\
\hline
($\alpha_{\bar q}=0.5$) \  & 300.7 & & & & \\
$\Delta u_v$  & &     0.357 $\pm$   0.093  &   0.000 $\pm$   0.001
                &     1.55  $\pm$   0.40   &   0.900 $\pm$   0.335 \\
$\Delta d_v$  & &  $-$0.512 $\pm$   0.098  &   0.159 $\pm$   0.227
                &     0.952 $\pm$   0.181  &   2.65  $\pm$   0.66  \\
$\Delta \bar q$ & & $-$9.50  $\pm$  10.07  &   0.5 (fixed)
                &  $-$0.980 $\pm$   0.060  &   0.0102 $\pm$   0.0394 \\
$\Delta g$    & &   148.5    $\pm$  7.4    &   2.11   $\pm$   0.26
                &  $-$1.0067 $\pm$  0.0005 &   0.051  $\pm$   0.337 \\
\hline
($\alpha_{\bar q}=1.0$) \  & 305.8 & & & & \\
$\Delta u_v$  & &     0.589   $\pm$   0.055  &   0.120 $\pm$   0.090
                &     0.632   $\pm$   0.059  &   1.62  $\pm$   0.27  \\
$\Delta d_v$  & &  $-$0.279   $\pm$   0.086  &   0.000 $\pm$   0.001
                &     2.58    $\pm$   0.80   &   1.32  $\pm$   0.31  \\
$\Delta \bar q$ & & $-$47.7   $\pm$   9.7    &   1.0 (fixed)
                &  $-$1.0065  $\pm$   0.0056 &   0.0204 $\pm$   0.0707 \\
$\Delta g$    & &   173.8     $\pm$  17.3    &   2.14  $\pm$   0.19
                &  $-$1.0058 $\pm$    0.0007 &   0.045 $\pm$   0.253  \\
\hline
($\alpha_{\bar q}=1.6$) \  & 323.5 & & & & \\
$\Delta u_v$  & &     1.356   $\pm$   0.132  &   0.335 $\pm$   0.120
                &  $-$0.477   $\pm$   0.046  &   0.313 $\pm$   0.209 \\
$\Delta d_v$  & &  $-$0.321   $\pm$   0.097  &   0.000 $\pm$   0.000
                &     2.12    $\pm$   0.64   &   2.21  $\pm$   0.17  \\
$\Delta \bar q$ & & $-$119.6  $\pm$  11.8    &   1.6 (fixed)
                &  $-$0.9976  $\pm$   0.0071 &   0.033 $\pm$   1.304 \\
$\Delta g$    & &   176.6     $\pm$   9.3    &   2.77  $\pm$   0.38
                &  $-$1.0057  $\pm$   0.0004 &   0.057 $\pm$   0.547 \\
\end{tabular}
\end{center}
\end{table}

\vspace{-0.5cm}
\begin{table}
\caption{Obtained first moments and spin contents at $Q^2=1$ GeV$^2$
         in the NLO analysis with fixed $\alpha_{\bar q}$.}
\label{T:1stmom-fix}
\begin{center}
\begin{tabular}{c|ccc}
distribution \    & $\alpha=0.5$  & $\alpha=1.0$  &  $\alpha=1.6$   \\
\hline
$\Delta \bar q$   &  $-$0.077  &  $-$0.057  &  $-$0.051    \\ 
$\Delta g$        &  0.550     &  0.533     &  0.294       \\ 
$\Delta\Sigma$    &  0.123     &  0.241     &  0.276
\end{tabular}
\end{center}
\end{table}

\vspace{-0.5cm}
\begin{table}
\caption{Obtained first moments.}
\label{T:1stmom}
\begin{center}
\begin{tabular}{c|cccc}
distribution \   & $Q^2$      &  LO                &  NLO-1    
                                                   &  NLO-2            \\
\hline
\                & 1 GeV$^2$  &                    &            &      \\
$\Delta u_v$     &            &    0.926 (fixed)   &    0.926 (fixed) 
                                                   &    0.926 (fixed)  \\ 
$\Delta d_v$     &            & $-$0.341 (fixed)   & $-$0.341 (fixed) 
                                                   & $-$0.341 (fixed)  \\ 
$\Delta \bar q$  &            & $-$0.064           & $-$0.089 
                                                   & $-$0.057          \\ 
$\Delta g$       &            &    0.831           &    0.532 
                                                   &    0.533          \\ 
$g_1^p$          &            &    0.144           &    0.110 
                                                   &    0.128          \\ 
$g_1^n$          &            & $-$0.067           & $-$0.069 
                                                   & $-$0.051          \\ 
$g_1^d$          &            &    0.036           &    0.019 
                                                   &    0.036          \\ 
\hline 
\                &  5 GeV$^2$ &                    &           &       \\
$\Delta u_v$     &            &    0.926           &    0.931
                                                   &    0.930          \\ 
$\Delta d_v$     &            & $-$0.344           & $-$0.343 
                                                   & $-$0.344          \\ 
$\Delta \bar q$  &            & $-$0.067           & $-$0.089 
                                                   & $-$0.059          \\ 
$\Delta g$       &            &    1.314           &    0.863
                                                   &    0.920          \\ 
$g_1^p$          &            &    0.143           &    0.118 
                                                   &    0.137          \\ 
$g_1^n$          &            & $-$0.068           & $-$0.075
                                                   & $-$0.056          \\
$g_1^d$          &            &    0.035           &    0.020 
                                                   &    0.038          \\ 
\hline
\                & 10 GeV$^2$ &                    &           &       \\
$\Delta u_v$     &            &    0.924           &    0.932  
                                                   &    0.931          \\ 
$\Delta d_v$     &            & $-$0.345           & $-$0.343 
                                                   & $-$0.345          \\ 
$\Delta \bar q$  &            & $-$0.068           & $-$0.089 
                                                   & $-$0.059          \\ 
$\Delta g$       &            &    1.524           &   0.999  
                                                   &   1.077           \\ 
$g_1^p$          &            &    0.143           &    0.120 
                                                   &    0.139          \\ 
$g_1^n$          &            & $-$0.068           & $-$0.076  
                                                   & $-$0.057          \\ 
$g_1^d$          &            &    0.035           &    0.021 
                                                   &    0.038          \\ 
\end{tabular}
\end{center}
\end{table}

\vfill\eject


\noindent

\vspace{-0.8cm}
\noindent

\vfill\eject

\vspace{+0.0cm}
\noindent

\vfill\eject

\vspace{+0.0cm}
\noindent

\vfill\eject

\vspace{+0.0cm}
\noindent

\vfill\eject

\ \\

\vspace{-2.8cm}
\noindent

\vfill\eject

\ \\

\vspace{-3.8cm}
\noindent

\vfill\eject

\vspace{+0.0cm}
\noindent

\vspace{0.0cm}
\noindent


\end{document}